\documentclass[%
 reprint,
 amsmath,amssymb,
 aps,
floatfix,
]{revtex4-2}

\usepackage{graphicx}% Include figure files
\usepackage{dcolumn}% Align table columns on decimal point
\usepackage{bm}% bold math

\usepackage[utf8]{inputenc}
\usepackage[T1]{fontenc}

\usepackage{setspace}
\usepackage{hyperref}
\usepackage[french]{varioref}
\usepackage{mathtools}
\usepackage{empheq}
\usepackage{amsmath}
\usepackage{amsfonts}
\usepackage{amssymb}
\usepackage{amsthm}
\usepackage{cancel}
\usepackage[dvipsnames]{xcolor}
\usepackage{epstopdf}
\usepackage{setspace}
\usepackage{tikz}
\usepackage{varwidth}
\usepackage{adjustbox}
\usepackage{array}
\usepackage{float}
\usepackage{tcolorbox}
\usepackage[normalem]{ulem}
\usepackage[tikz]{mdframed}

\mdfdefinestyle{MyFrame}{%
    linecolor=black!50!gray,
    linewidth=2pt,
    roundcorner=10pt,
    innertopmargin=0\baselineskip,
    innerbottommargin=\baselineskip,
    innerrightmargin=20pt,
    innerleftmargin=20pt,
    backgroundcolor=gray!25!white}

\usetikzlibrary{shapes.geometric, arrows}

\tikzstyle{startstop} = [rectangle, rounded corners, minimum width=3cm, minimum height=1cm,text centered, draw=black, fill=red!30]
\tikzstyle{io} = [trapezium, trapezium left angle=70, trapezium right angle=110, minimum width=3cm, minimum height=1cm, text centered,
	execute at begin node={\begin{varwidth}{3.5cm} \begin{center}},
   execute at end node={\end{center} \end{varwidth}}, draw=black, fill=blue!30]
\tikzstyle{process} = [rectangle, minimum width=3cm, minimum height=1cm, text centered, 
	execute at begin node={\begin{varwidth}{5cm} \begin{center}},
   execute at end node={\end{center} \end{varwidth}}, draw=black, fill=orange!30]
\tikzstyle{decision} = [diamond, minimum width=3cm, minimum height=1cm, text centered, 
	execute at begin node={\begin{varwidth}{2cm} \begin{center}},
   execute at end node={\end{center} \end{varwidth}}, draw=black, fill=green!30]
\tikzstyle{arrow} = [thick,->,>=stealth]
\tikzstyle{holder} = [rectangle, minimum width=11cm, minimum height=7cm, text centered, draw=black, fill=pink!100]

\newtheorem*{theorem*}{Théorème}

\date{\today}

\begin{document}
\title{Gravitational influence of high power laser pulses}
\author{Paul Lageyre}
\email{paul.lageyre@net-c.fr}
\author{Emmanuel d'Humières}
\author{Xavier Ribeyre}
\affiliation{CEntre Laser Intenses et Applications, Univ. Bordeaux-CNRS-CEA, UMR 5107 Talence 33405, France}
\begin{abstract}
    The study of the generation of metric perturbation in the laboratory presents an opportunity to observe and understand more easily the mechanisms at work in gravitation. The present study will focus on the metric perturbation generated by a light pulse, as it could be generated by a current ultra-high power laser. Although of very small magnitude, the potential thus generated has advantages over that generated by mass acceleration, such as the absence of noise due to non uniform acceleration or the ability to scale up the experiment. It is indeed easier to scale up an electromagnetic oscillation compared to a mechanical oscillator, which must either be made with a large accelerated mass or a lot of small masses, all in sync, which acceleration must furthermore be quadripolar. Generation of metric deformation by laser could therefore prove useful in the long-term establishment of a laboratory experiment for the generation and detection of gravitational waves.
\end{abstract}
\maketitle
\renewcommand{\labelitemi}{$\bullet$}

\section{Introduction}
The detection of metric perturbations limited in time is for the scientific community a chance to better understand the subtleties of the gravitational force and to confirm, or complete, its description by general relativity, theorized by Einstein between 1907 and 1915. Propagative transverse deformations of the space-time metric, the gravitational waves, are themselves theorized by \citet{einstein_naherungsweise_1916}.

The first gravitational waves are detected indirectly for the first time in 1974 thanks to the observations by Hulse and Taylor of a binary system of a pulsar and a neutron star, whose orbital period was affected by the radiation of a gravitational wave \citep{hulse_discovery_1975}.
As for the first direct detection, it took place at the end of 2015 \citep{abbott_observation_2016}. It was made possible by the construction of giant interferometers such as LIGO, whose very high sensitivity has allowed the observation of changes in the length of its arms of the order of $\Delta L/L=10^{-22}$ for frequencies on the $\sim 100$ Hz range.
This new observation has rekindled the scientific community's interest in gravitational waves and what they can tell us about the physics of gravitation.

During the century that separates the prediction of gravitational waves from their observation, many methods of generating gravitational waves and metric perturbations in the laboratory have been studied. Indeed, such a source would have the advantage of being more reproducible and adaptable to the desired observations than the observation of very intense astrophysical phenomena, even though they are more commonly observed nowadays \cite{abbott_gw170817_2017,abbott_gwtc-1_2019,abbott_gwtc-2_2021,the_ligo_scientific_collaboration_gwtc-3_2021}.
Two approaches are favored for the generation of a large enough to be detected time varying gravitational potential in the laboratory. They both require powerful sources to generate a significant deformation. \\
The first is the explosive acceleration of a quantity of mass, which would generate a deformation of space-time through the quadrupolar acceleration of mass. This deformation can be conceived by the means of a laser striking a target \cite{ribeyre_high_2012,gelfer_gravitational_2016,kadlecova_gravitational_2017} or in a more extreme manner by the explosion of a thermonuclear bomb \citep{chapline_gravitational-radiation_1974}.
The second is the generation of a metric deformation by an electromagnetic wave, whose coherence is an advantage in the generation of an important deformation \citep{grishchuk_electromagnetic_2003}.

Indeed, the generation of a powerful electromagnetic wave is a process that, in lasers, forces the coherence of the electromagnetic oscillation. Such wave thus has a clear direction of propagation at any point of its existence, and a determined spectrum. Unlike its purely electromagnetic counterpart, mass acceleration by laser is a chaotic process, since it involves the deposit of an important amount of energy on a material, which then causes the mass acceleration. Such acceleration is thus not only quadripolar and a lot of energy can be lost not contributing to the generation of a detectable metric perturbation, either through dipolar acceleration or material heating.

The observation in astrophysics of the deviation of the ray of light by Dyson and Eddington \citep{dyson_determination_1920} constitutes one of the first important tests of general relativity. This test confirms the influence the variation of the gravitational potential has on the trajectory of light in space. From the point of view of general relativity, it confirms that the deformation of space-time generated by a massive object (in this case the Sun) changes the trajectory of light which then follows the geodesic in this new space-time. From a Newtonian point of view, we can say that the Sun exerts an influence on light through the gravitational force, which implies, by reciprocity of action, that light has itself an influence on gravity. This analogy suggests that light must generate locally its own gravitational field, that is to say a deformation of space-time. From the relativistic point of view, light carries energy, and must therefore generate a gravitational potential.
The first to be interested in the gravitational potential produced by light were Tolman, Ehrenfest and Podolsky \citep{tolman_gravitational_1931} who studied the deformation generated by an infinitesimally thin ray of light on the surrounding space-time. Gravitational generation by light is an old problem and has been considered by several authors \cite{gertsenshtein_wave_1962,kolosnitsyn_gravitational_2015, scully_general-relativistic_1979, grishchuk_update_2003}. This study of the metric deformation generated by a beam of light of zero spatial extension has more recently been taken up and completed by \citet{ratzel_gravitational_2016}, but the absence of a spatial extension for this beam of light prevents the complete study of metric deformation.

In this paper, we shall consider the generation of a metric perturbation inside a light beam, where it should be the most important. Moreover, we will present an exact solution of the Einstein equations on the light cylinder axis and consider both in the laboratory and astrophysical situations.
In section 2, we will summarize linearized general relativity for small deformations of space-time, and apply it to the case of an electromagnetic wave. In section 3, we will present the analytical method used to solve this problem, and, in section 4, we will check the compatibility of our solution with the already known case of the linearized \citet{schwarzschild_uber_1916} metric. In section 5, we will apply this calculation method to a simple model of a light pulse of circular polarization: the cylinder of light, of constant energy density and moving at the speed of light in vacuum $c$. In section 6, we will analyze the results obtained by keeping in mind the characteristics of the laser sources that could be used in a laboratory experiment. We will expand this study to the case of a linearly polarized light pulse by first studying the oscillatory term thus introduced alone in section 7, before analyzing the whole solution in section 8.  In section 9, we will compare our results with those obtained by \citet{ribeyre_high_2012}, \citet{gelfer_gravitational_2016} and \citet{kadlecova_gravitational_2017} in the case of a massive source. We will also mention the future developments of this paper, as well as give some insight on the possibility of detection of the gravitational phenomenon studied here, both in laboratory and in astrophysical settings. Section 10 will in the end highlight the most important points pulled from this study.
%%%%%%%%%%%%%%%%%%%%%%%%%%%%%%%%%%%%%%%%%%%%%%%%%%%%%%%%%%%%%%%
\section{Metric deformation generated by an electromagnetic field}
We assume the existence of a plane space-time described by the Minkowsky metric $\eta_{\mu\nu}= diag(-1,1,1,1)$, corresponding to the approximation of a Galilean frame of reference, where the local distance traveled on the interval $ds$ by an object can be described as:
\begin{equation}
    ds^2=\eta_{\mu\nu} dx^{\mu}dx^{\nu}= -c^2 dt^2 + dx^2 + dy^2 + dz^2
\end{equation}
By analogy, the metric of a deformed spacetime can be written as $g_{\mu\nu}$, which can be written for a spacetime where the metric perturbation with respect to the planar spacetime is negligible as:
\begin{align}
    g_{\mu\nu}=\eta_{\mu\nu} + h_{\mu\nu}
\end{align}
Where $h_{\mu\nu} \ll 1$ is the perturbation in the $g_{\mu\nu}$ metric in respect to the Minkowsky metric $\eta_{\mu\nu}$.\\
In such a framework, Einstein's equations \citep{landau_classical_2009}, which relate the structure of the space-time studied with the stress-energy tensor of the various elements involved, can be written as:
\begin{align}
    G_{\mu\nu}=\chi T_{\mu\nu}
\end{align}
Where $G_{\mu\nu}$ is the Einstein's tensor which describes the local space-time curvature. $T_{\mu\nu}$ is the stress-energy tensor, and $\chi=8\pi G/c^4$ is Einstein's constant which depends on the gravitational constant $G$ and on the speed of light in vacuum $c$. Applying the Lorentz gauge condition, we get:
\begin{equation}
    \partial^{\mu}h_{\mu\nu}=0
\end{equation}
Which gives us the linearized Einstein equations:
\begin{equation}
    \square h_{\mu\nu}=-2\chi T_{\mu\nu}
    \label{eq:gravwave}
\end{equation}
Which are none other than d'Alembert equations, i.e wave equations. The stress-energy tensor can be rewritten as the sum of matter's stress-energy tensor $T_{\mu\nu}^{mat}$ and electromagnetic field's stress-energy tensor $T_{\mu\nu}^{em}$.

We are interested here in the generation of a metric perturbation by a light pulse, so we will take:
\begin{equation}
    T_{\mu\nu}^{mat}=0 \text{  et  } T_{\mu\nu}^{em}=-\sigma_{\mu\nu}
    \label{eq:EImpEm&Mat}
\end{equation}
Where $\sigma_{\mu\nu}$ is the Maxwell stress-energy tensor in $4\times 4$ dimension, symmetrical such that in SI notation \citep{boudenot_electromagnetisme_1989}:
\begin{align}
    &-\sigma_{00}=\epsilon_0\frac{E^2+c^2B^2}{2} \\ &\text{the energy density.} \notag\\
    &-\sigma_{0k}=-\sigma_{k0}=\frac{\epsilon_{ijk}E_i B_j}{c\mu_0} \\ &\text{the energy flux in the direction }x^k \text{ (} k=1,2,3 \text{)}.\notag\\
    &\sigma_{ij}=\epsilon_0 E_i E_j + \dfrac{B_iB_j}{\mu_0} - \delta_{ij}\epsilon_0\frac{E^2+c^2B^2}{2} \\ &\text{the 3D Maxwell constrain tensor.} \notag
\end{align}
With $i,j,k = 1,2,3$ the spatial components' indices and $\delta_{ij}$ the Kronecker symbol.

Armed with equations \eqref{eq:gravwave} and \eqref{eq:EImpEm&Mat} which predict a deformation of the space-time in the presence of an electromagnetic field, we can study the metric perturbation generated by a light pulse propagating in an initially flat space-time.

Let us take the example of a plane progressive wave of linear polarisation traveling towards positive $z$ axis:
\begin{gather}
    {\bf{E}}= E {\bf{e_x}} \text{  and }
    {\bf{B}}= B {\bf{e_y}}\\
    E=cB \notag
\end{gather}
The tensor $\sigma_{\mu\nu}$ can then be written as:
\begin{equation}\label{eq:Maxwell_T}
    \sigma_{\mu\nu}=-\epsilon_0 \begin{pmatrix}
    E^2 & 0 & 0 & E^2\\
    0 & 0 & 0 & 0\\
    0 & 0 & 0 & 0\\
    E^2 & 0 & 0 & E^2
    \end{pmatrix}
\end{equation}
We are thus expecting a metric perturbation of space-time $h_{\mu\nu}$ propagating longitudinally according to the direction of the source progressive electromagnetic wave:
\begin{gather}
    h_{00}=h_{03}=h_{30}=h_{33} \equiv \phi\\
    h_{\mu\nu}=0 \text{  for } \mu \neq 0,3 \text{ or } \nu \neq 0,3
\end{gather}
This deformation profile is to be contrasted with that of the commonly studied gravitational waves, which propagate at long distances only in deformations transverse to their propagation \citep{maggiore_gravitational_2008}.

Let us clarify the source $E^2$ present in the d'Alembert equation \eqref{eq:gravwave}:
\begin{equation}
    E=E_0 \cos(k(z-ct))
\end{equation}
Where $E_0$ is the amplitude of the wave's electric field, and $k=2\pi/\lambda$ the wave vector of the electromagnetic wave propagating towards positive $z$.\\
Hence with equation \eqref{eq:Maxwell_T}:
\begin{align}
\square \phi&=-2\chi ~\epsilon_0 E_0^2 \cos^2 (k(z-ct))\\    
&=-\chi ~\epsilon_0 E_0^2 \left(1+\cos(2k(z-ct))\right) \label{eq:hSource}
\end{align}
The expression \eqref{eq:hSource} seems trivial but shows explicitly that we can decompose the source term into its constant part and its oscillating part, and solve these two parts separately since the d'Alembertian is a linear operator.
We here interest ourselves in solving the linearized Einstein equation \eqref{eq:hSource} first for the non-oscillatory part, then for the oscillatory part of the source term.

The constant alone in the electromagnetic stress-energy tensor also happens to correspond to the energy density of a circularly polarized electromagnetic wave. The following study thus provides a complete solution for the longitudinal metric deformation generated in this specific physical case.\\
The Einstein equation for the study of a light pulse of circular polarisation can be written as:
\begin{equation}
    -\square \phi = 2\chi S
    \label{eq:hSourceCte}
\end{equation}
where $S$ is the local source function link to the electromagnetic stress-energy tensor. For the case of a circularly polarized light, $S=\epsilon_0 E_0^2/2$ is the mean energy density of the light pulse.\\ In such case, we will write the $\phi$ thus studied as $\phi_0$, as it describes a non-oscillating solution. In a similar way, we will note $\phi_k$ the metric perturbation caused by an oscillating source term of wave vector $k$. For the case of linearly polarized light as described in equation \eqref{eq:hSource}, we can thus write: $\phi = \phi_0 + \phi_{2k}$.\\
We will write equations with $\phi$ as long as they stay true for any studied source term.

Please note that when the source term $S$ is a constant, equation \eqref{eq:hSourceCte} devolves into the classical gravitational Poisson equation for a static gravitational potential as the D'Alembertian operator $\square$ becomes in absence of time dependance the Laplacian operator $\Delta$.
We investigate the solution of Equation \eqref{eq:hSourceCte} for a light pulse of finite spatial and temporal extension, such as it could be emitted by a laser system. This case is, as such, time dependant and thus cannot be reduced to the gravitational Poisson equation. This uniform localized source term allows to propagate gravitational influence like a "soliton".
This light pulse will be modeled by a cylinder of constant energy density of length $L$ and radius $R$ moving at the speed of light $c$ in the direction of positive $z$. We propose to study an intermediate case, that of the static cylinder, in order to illustrate in a simpler case the methods used and to verify the compatibility of our method of resolution with the well known solution of the metric of \citet{schwarzschild_uber_1916} for a stationary isolated mass.
%%%%%%%%%%%%%%%%%%%%%%%%%%%%%%%%%%%%%%%%%%%%%%%%%%%%%%%%%%%%%%%%%
\section{Resolution in the simple case of a static cylinder} \label{sec:CylImmob}
Equation \eqref{eq:hSourceCte} is a partial differential equation on 4 dimensions. The solution of the homogeneous equation is the set of functions of $\rho-ct$ or $\rho+ct$, where $\boldsymbol{\rho}$ is the position vector of a point in space only. Since the particular solution of this equation is not apparent, we have to use Green's function $G_\square$ of the d'Alembertian operator to determine a solution.
\begin{gather}
    \square G_\square ({\bf{x}} - {\bf{x'}})=\delta({\bf{x}} - {\bf{x'}})\\
    \rm and \notag\\
    \phi({\bf{x}})= -2\chi \int_{\mathbb{R} ^4} d{\bf{x'}}  G_\square ({\bf{x}} - {\bf{x'}}) S({\bf{x'}}) \label{eq:GreenBaseSol}
\end{gather}
Where ${\bf{x}} \equiv (ct,{\boldsymbol{\rho}})$ is the observer's time space position vector and ${\bf{x'}}$ another vector describing a position in spacetime.
Thankfully, the d'Alembertian's Green function is well-known \citep{jackson_classical_1999} and can be written in spherical coordinates for a retarded potential as:
\begin{equation}
    G_\square \left(c(t-t'),|{\boldsymbol{\rho}}-{\boldsymbol{\rho'}}|\right) =-\dfrac{\delta(c(t-t')-|{\boldsymbol{\rho}}-{\boldsymbol{\rho'}}|)}{4\pi|{\boldsymbol{\rho}}-{\boldsymbol{\rho'}}|} 
    \label{GreenExpr}
\end{equation}
We can then focus on solving the successive integrals of Equation \eqref{eq:GreenBaseSol}. As the following calculation highlights some techniques, we will pass by the example of a cylinder of constant energy density to illustrate them.

\begin{figure}
\begin{center}
\begin{minipage}{.6\linewidth}
\includegraphics[width=\linewidth]{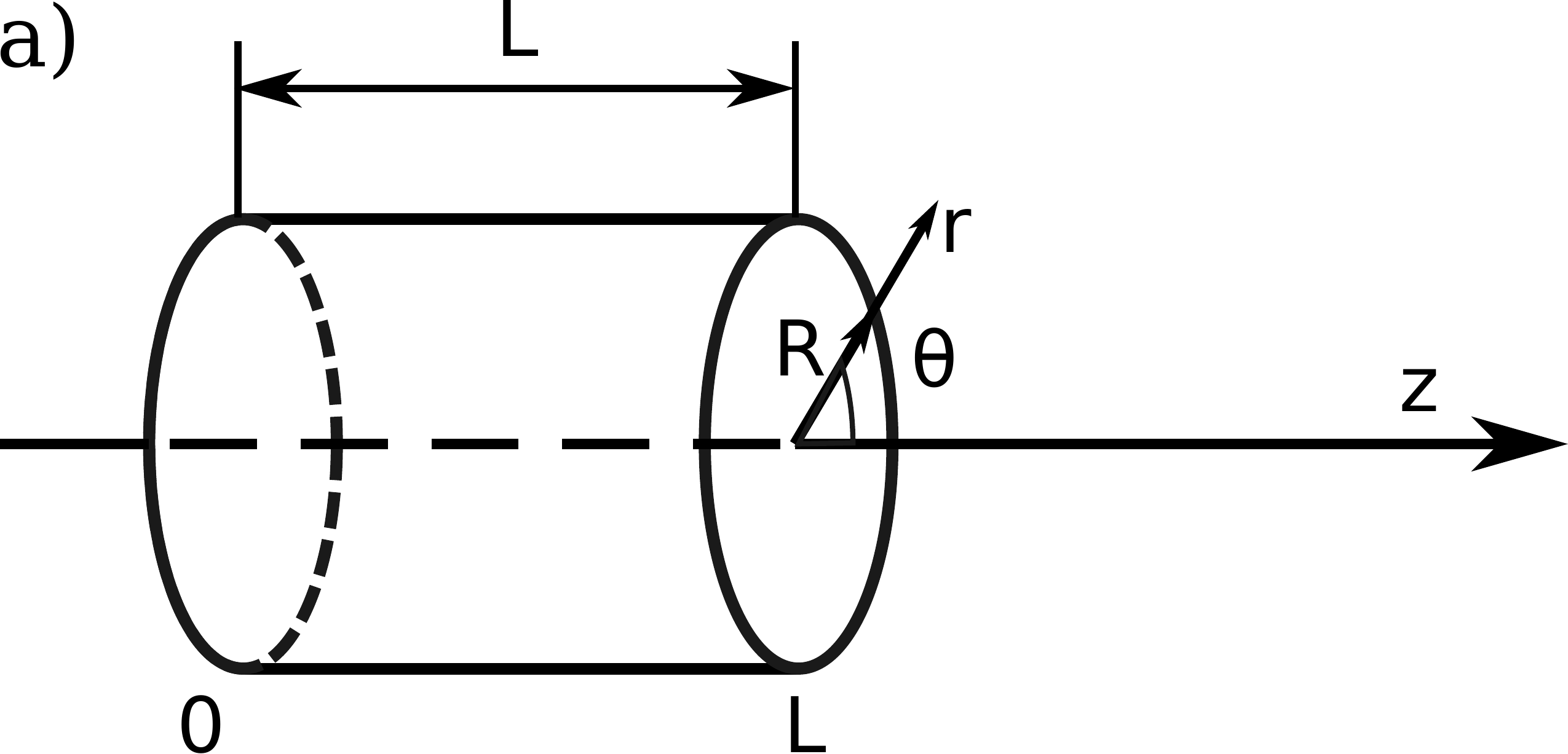}
\end{minipage}
%%%%%%%%%%%%%%%%%%%%%%%%%%%%%%%%%%%%%%%%%%%%%%%%%%%%%%%%%%%%%%
\vspace{5pt}
\begin{minipage}{\linewidth}
\includegraphics[width=\linewidth]{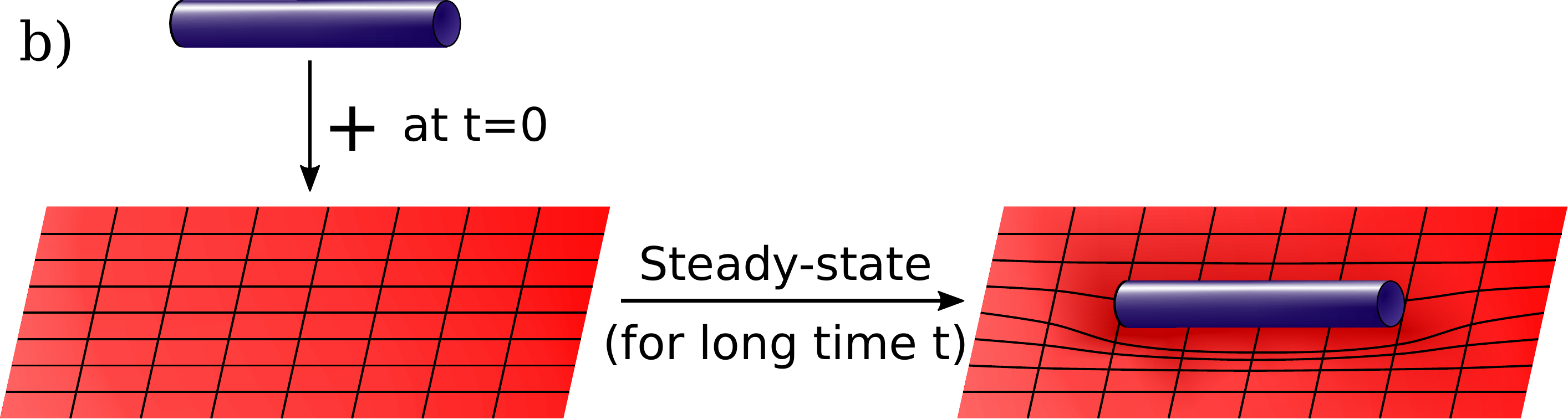}
\end{minipage}
\caption{a) Schematic representation of the studied static cylinder of energy density and b) Presentation of the thought experiment modeling the establishment of the gravitational potential of said cylinder at $t=0$ and $t\gg0$. \label{CylindreStatBisCropped}}
\end{center}
\end{figure}

We represent the theoretical physical situation in Figure~\ref{CylindreStatBisCropped}. We will here take a cylinder of radius $R$ and length $L$ so that the axis $Oz$ is the axis of symmetry of the cylinder. The cylinder starts on this axis at $z=0$ and ends at $z=L$. This cylinder appears at time $t=0$ in a previously planar space-time. It generates a deformation which must be, at long time and far from the source, the one described by Newtonian physics for an object of equivalent mass density $\rho_m={\cal A}/c^2$. For this cylinder of homogeneous energy density, we have the following:
\begin{equation}
 S(ct,z,r)= {\cal A} H(ct)H(z)H(L-z)H(R-r).
 \label{eq:SrcStat}
\end{equation}
Where $\cal{A}$ is the constant energy density inside the cylinder, and $H(z)$ is the Heaviside function. We will note the metric perturbation thus generated as $\phi_c$. It is interesting to remark that the light cylinder appears in the space-time at t=0 as it is the case in Ref. \cite{ratzel_gravitational_2016}, this allow to study the non-stationary nature of the solution.
We can then rewrite Equation \eqref{eq:GreenBaseSol} in cylindrical coordinates. Taking into account that, in Equation \eqref{GreenExpr} we have $|{{\bf{\rho}}-\bf{\rho'}}|=\sqrt{(z-z')^2+(r^2+r'^2-2rr'cos\theta')}$, we get:
\begin{widetext}
\begin{align}
    \phi(ct,z,r,\theta)= \dfrac{2\chi}{4\pi} \int^{\infty}_{-\infty} c dt' \int^{\infty}_{-\infty} dz' \int^{\infty}_{0} r' dr' \int^{2\pi}_{0} d\theta ' & \, S\left(ct',z',r',\theta '\right) \dfrac{\delta(c(t-t')-|{{\bf{\rho}}-\bf{\rho'}}|)}{|{{\bf{\rho}}-\bf{\rho'}}|} \\
    \phi(ct,z,r,\theta)= \dfrac{\chi}{2\pi} \int^{\infty}_{-\infty} dz' \int^{\infty}_{0} dr' \int^{2\pi}_{0} d\theta ' \, S & \left(ct-\sqrt{(z-z')^2+({\bf{r}}-{\bf{r'}})^2},z',r',\theta '\right) \label{SolGreen0} \\
  &\times \dfrac{r'}{\sqrt{(z-z')^2+(r^2+r'^2-2rr'cos\theta')}}\notag
\end{align}
\end{widetext}
For the sake of simplicity, we only calculate the gravitational influence of the cylinder on the $Oz$ axis, the axis of symmetry of the cylinder. We thus have $r=0$, which leads us to note $\phi(ct,z,r,\theta)=\phi(ct,z)$. By introducing Equation \eqref{eq:SrcStat}, we then obtain:

\begin{align}
& \phi_c(ct,z) = \dfrac{\chi {\cal{A}}}{2\pi}\int^{\infty}_{-\infty} dz' H(z') H(L-z') \times\\
&\int^{\infty}_{0} H(ct-\sqrt{r'^2 + (z'-z)^2}) H(R-r') \dfrac{r' dr'}{\sqrt{(z-z')^2+r'^2}}\notag
\end{align}

Heaviside functions exhibit conditions for integration. Among these conditions, the condition to linearize with respect to $r'$ to perform the integration is:
\begin{equation}
    ct-\sqrt{r'^2 + (z'-z)^2} \geq 0 \Rightarrow \left\lbrace \begin{array}{l}
    r'^2   \leq   c^2t^2 - (z'-z)^2\\
    c^2t^2 - (z'-z)^2 \geq 0
    \end{array}\right.
\end{equation}
By studying the different conditions, we reach the expression:
\begin{widetext}
\begin{equation}
\begin{split}
\phi_c(ct,z,0)= {\cal{A}} \chi & \int^{\infty}_{-\infty}  dz' H(z')H(L-z') H(z'-(z-ct)) H((z+ct)-z')  \overbrace{\left[-\sqrt{(z-z')^2} \right.}^\text{\small{lower integration bound}} \\
 & \underbrace{\left. + H\left(\sqrt{c^2t^2-(z-z')^2}-R\right)~ \sqrt{(z-z')^2+R^2} + H\left(R-\sqrt{c^2t^2-(z-z')^2}\right)~ ct \right]}_\text{\small{upper integration bound}}
\end{split}
\label{CylStat1stInt}
\end{equation}
\end{widetext}
The integration can be carried out "by hand", but the analytical solution is difficult to interpret. It is therefore preferable to linearize the conditions on $z'$, find the functions' primitives, and provide them to a Python program \footnote{The code used to visualize the solutions is accessible in Paul Lageyre's github repository: \url{https://github.com/Paul-Lageyre/2021_plotting_code}} that will compare the different terms and conditions to plot the solution of Equation \eqref{CylStat1stInt}.
%%%%%%%%%%%%%%%%%%%%%%%%%%%%%%%%%%%%%%%%%%%%%%%%%%%%%%%%%%%%%%%%%%%%%%%%%%%%
\section{Visualisation and study of the analytical solution}
\begin{figure}
\begin{center}
\begin{minipage}{\linewidth}
\begin{flushright}
\includegraphics[width=.9\linewidth]{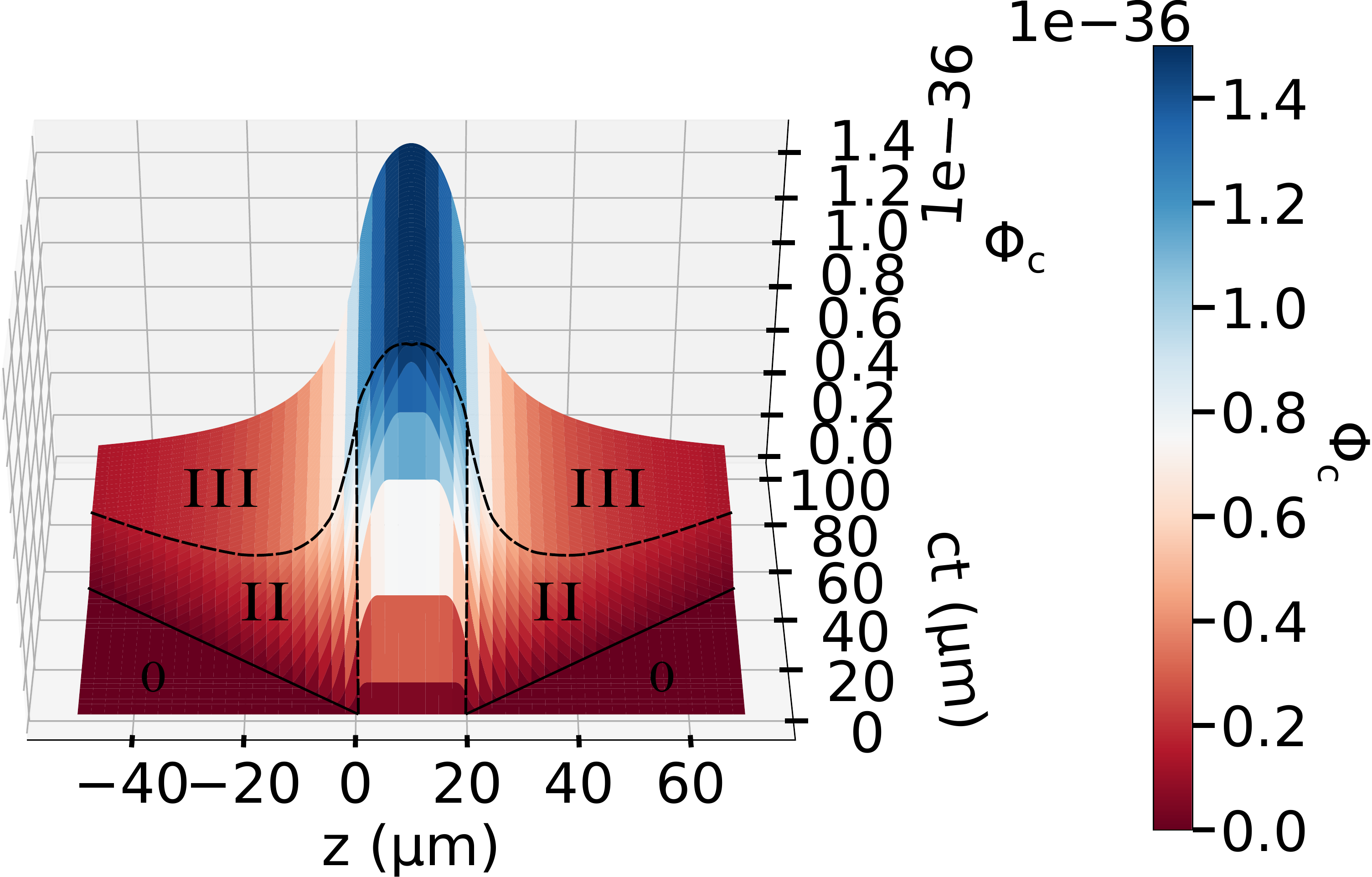}
\end{flushright}
\end{minipage}
%%%%%%%%%%%%%%%%%%%%%%%%%%%%%%%%%%%%%%%%%%%%
\begin{minipage}{\linewidth}
\includegraphics[width=\linewidth]{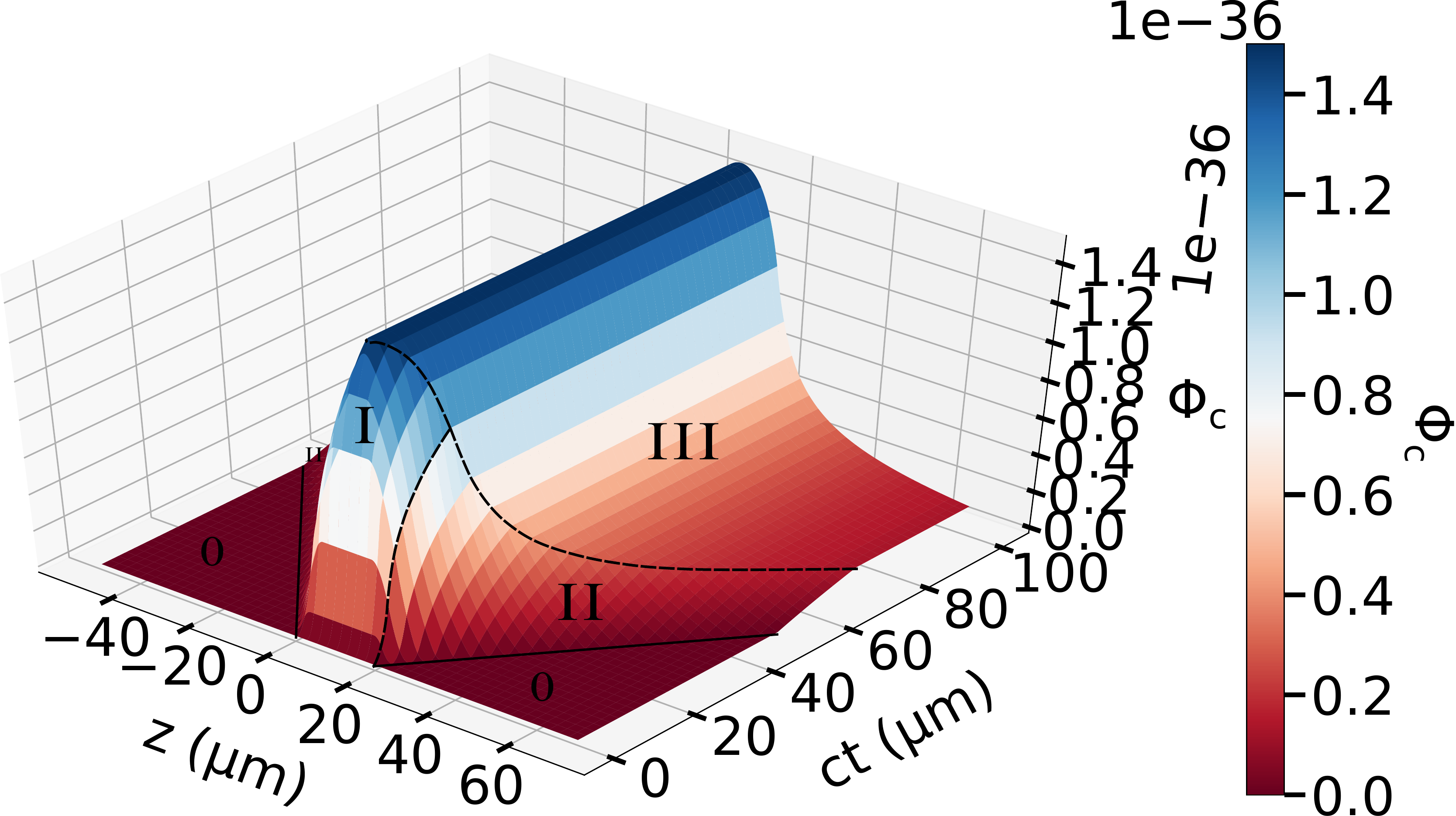}
\end{minipage}
\caption{3D representation according to two angles {\large (}a) "head-on" and b) "front-on"{\large )} of the solution on the plane $(O,z,ct)$ for a laser intensity $I=10^{22} \, \rm W/cm^2$ and dimensions of the light cylinder $L=20~\rm \mu m,~ R=~5~\rm \mu m$. The numbers indicate the solution's regimes. Zone I: the transient regime inside the source, zone II: the transient regime outside the source, zone III: the permanent regime, zone 0: unperturbed spacetime. \label{3DStat}}
\end{center}
\end{figure}

In order to model what a current petawatt-class power laser could produce \citep{danson_petawatt_2019}, we take the following values: ${\cal A} \times c=I= 10^{22} \,\rm W/cm^2,~ E=525 \,\rm J$
which correspond to the following dimensions of the static cylinder: $L=20~\rm \mu m,~ R=5~\rm \mu m$. The solution is shown in Figure \ref{3DStat}.
This visualization highlights the establishment (zone I and II) of the stationary solution (zone III) for the metric deformation of a static cylinder of constant energy density. We thus have the exact solution for an out of equilibrium space-time, a point which will be very useful for the study of the cylinder moving at the speed of light.

This transient regime leads to a stationary deformation in time which corresponds to a gravitational potential. The stationary potential thus established in zone III is the gravitational potential of the cylinder of stationary energy density. This potential is the strict analogue of the potential which would be generated by a massive object of cylindrical shape along its axis.  This potential behaves at long distance as $1/z$, i.e. like the gravitational potential of a massive object at long distance. 
Indeed let us take for example $z$ large and negative, in order to have a clearly readable development of the gravitational potential far from the source. We obtain by Taylor expansion:
\begin{gather}
\phi_c(z,L,R)/\chi \cal{A} \notag\\
= \notag \\
-L(L-2z) + (L-z)\sqrt{R^2 + (z-L)^2}+ \notag \\
+R^2 {\rm arcsh}\left( \dfrac{L-z}{R} \right) + z\sqrt{R^2 + z^2} - R^2 {\rm arcsh}\left( \dfrac{-z}{R} \right) \notag \\
\approx\notag \\
\dfrac{LR^2}{-z}
\end{gather}
Be careful, here the potential thus generated is positive, for $z$ is negative. We therefore find a potential similar to the $1/r$ gravitational potential found in the linearized \citet{schwarzschild_uber_1916}'s metric model.

\begin{figure}
\begin{center}
\includegraphics[width=\linewidth]{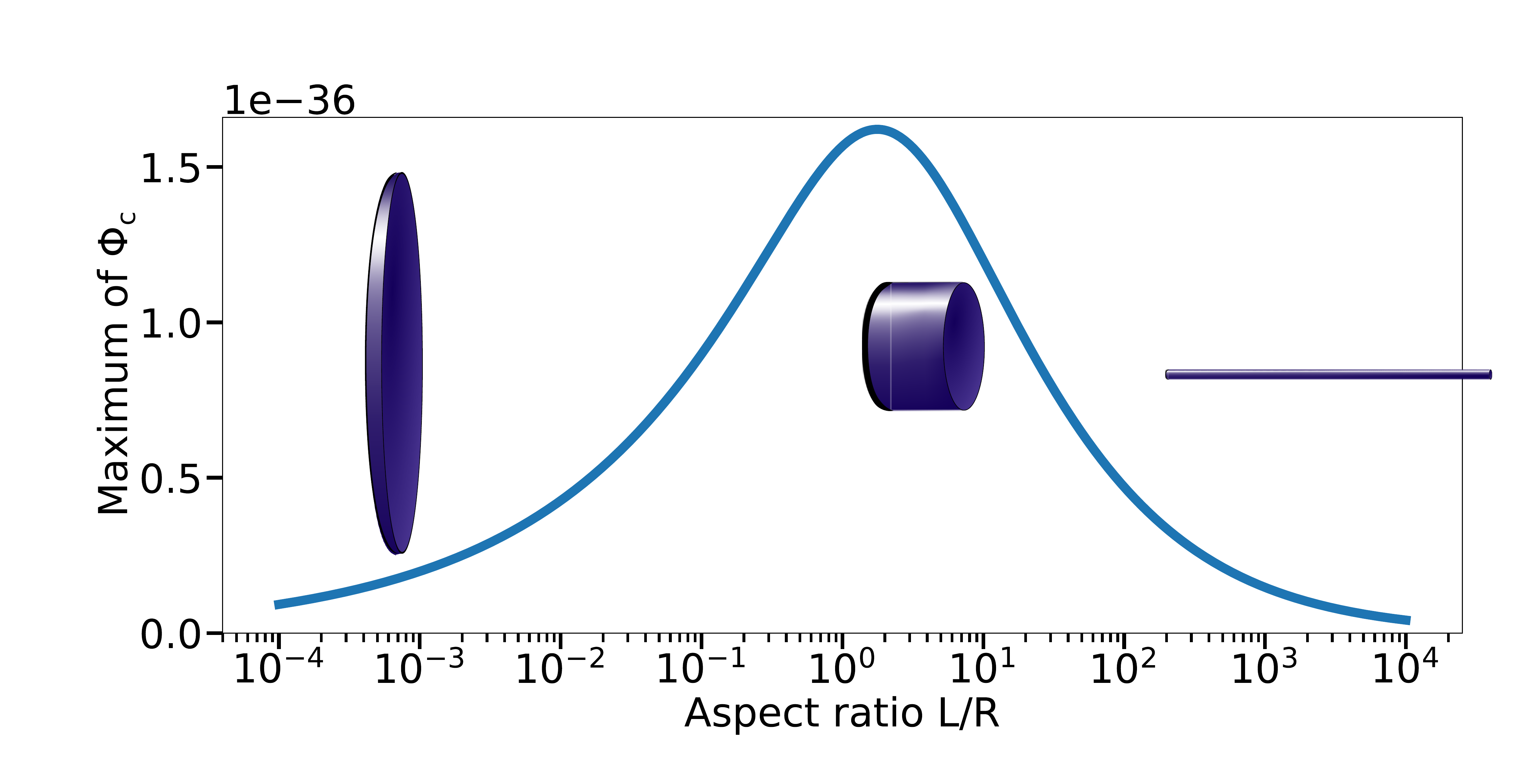}
\caption{Maximum amplitude of the metric deformation $\phi$ generated by a cylinder of constant energy density ${\cal{A}}= 3.3 \times 10^{11} \, \rm W/cm^3$ and constant energy $E=525 \, \rm J$ as a function of the aspect ratio $L/R$ of the cylinder. The 3 cylinders represent the general shape of the cylinder for the different aspect ratio domains in the figure.}\label{MaxStat}
\end{center}
\end{figure}

If we study cylinders of different aspect ratios $L/R$ at constant $I$ and $E$, it appears that the maximum metric deformation on the $Oz$ axis varies. This maximum is plotted as a function of the aspect ratio Figure \ref{MaxStat}. The abscissa of the maximum of metric deformation as a function of the aspect ratio Figure \ref{MaxStat} cannot be explained only by a geometrical argument, we will stop there in the study of the case of a static cylinder, which already gives us a reliable result. In our example the maximum of the perturbation $\phi$ is $8\times 10^{-37}$ for $L/R\sim1.77$ with $I=10^{22} \, \rm W/cm^2$ and $E=525 \, \rm J$.
Armed with the observations made and the tools developed in this section, we will extend this study to the more realistic case of a cylinder of light moving at the speed of light $c$.
%%%%%%%%%%%%%%%%%%%%%%%%%%%%%%%%%%%%%%%%%%%%%%%%%%%%%%%%%%%%%%%%
\section{Light pulse model: cylinder moving at speed \textit{c}}
\begin{figure}
\begin{center}
\begin{minipage}{.6\linewidth}
\includegraphics[width=\linewidth]{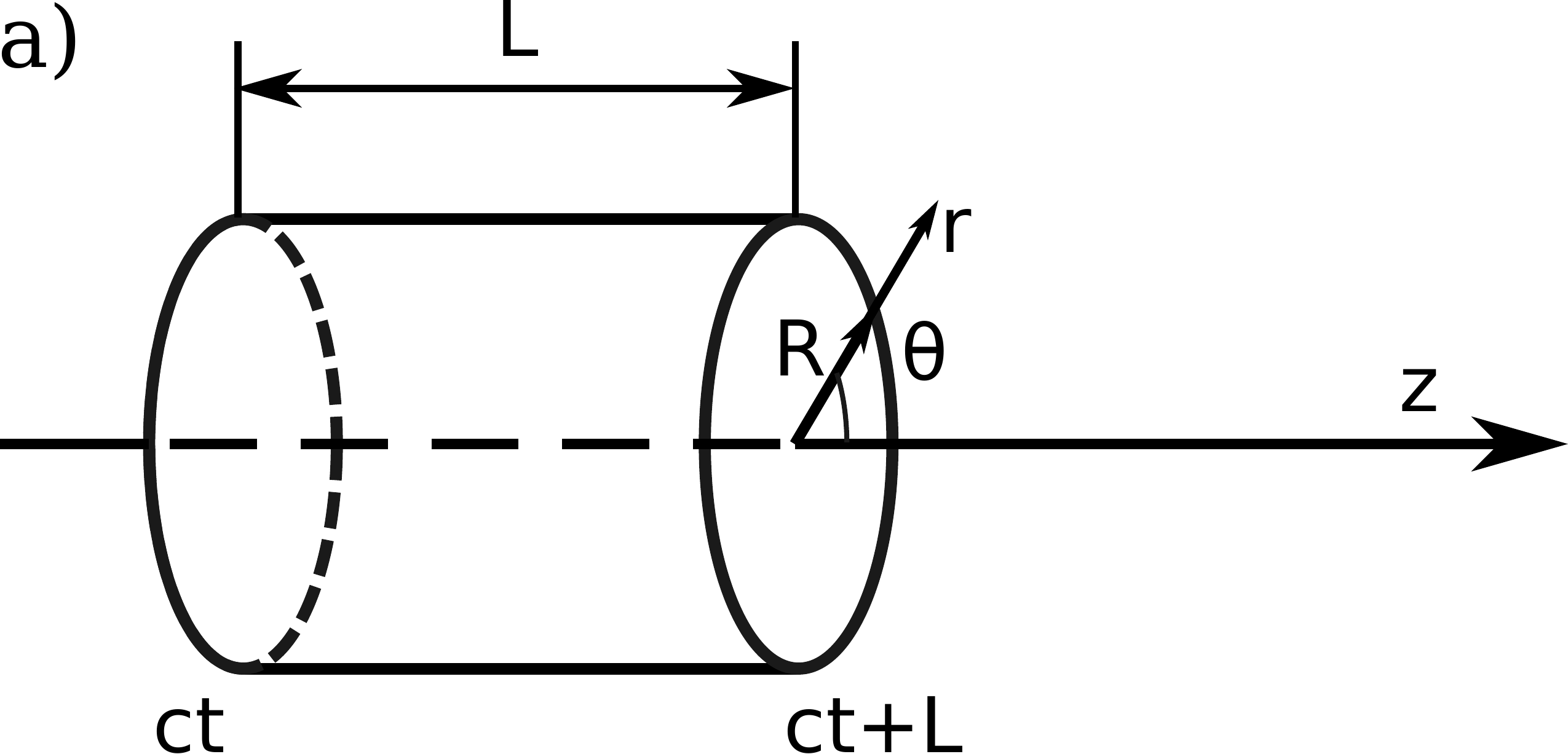}
\end{minipage}
\vspace{8pt}
\begin{minipage}{\linewidth}
\includegraphics[width=\linewidth]{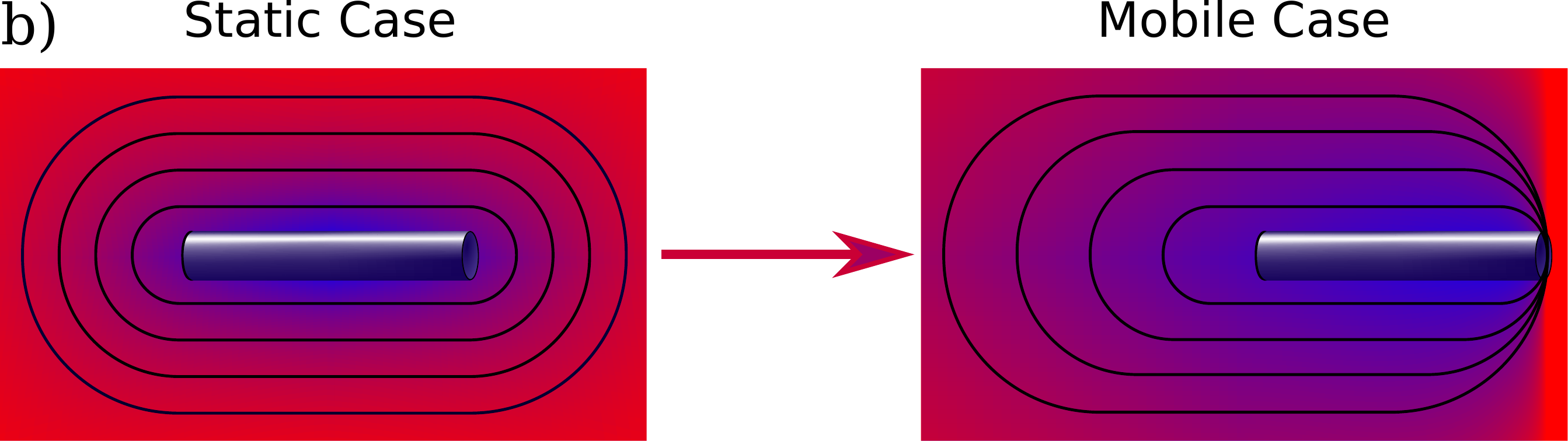}
\end{minipage}
%\begin{tikzpicture}[node distance=1.8cm]
%\node (start) [holder] {\Large{\textsc{CylindreMob (Figure non finie)}}};
%\end{tikzpicture}
\caption{a) Schematic representation of the cylinder of uniform energy density moving at the speed of light and b) Illustration of the expected differences between the metric deformation generated by the static cylinder previously studied and that generated by a light pulse of the same cylindrical shape.\label{CylindreMob}}
\end{center}
\end{figure}

We repeat here the method presented previously for a static cylinder. We now consider the cylinder of light moving at the speed of light $c$. This problem is illustrated in Figure \ref{CylindreMob}. This cylinder moves in the direction of positive $z$ in such a way that the cylinder of light is located at time $t$ between $z=ct$ and $z=ct+L$. It appears at $t=0$ in a planar space-time. The energy density in spacetime can therefore be written as:

\begin{equation} \label{eq:SrcCte}
\begin{split}
    S & (ct,z,r,\theta)=\\
    & {\cal{A}} H(ct)H(R-r)H(r) H(z-ct)H(L-(z-ct))
\end{split}
\end{equation}
A first rough analogy would be to compare this case of a longitudinal deformation propagating at the same speed as the object that generates it with the already known case of an object moving at the speed of sound in a medium. This analogy highlights the particularity of a case where the source and the perturbation propagate at exactly the same speed as compared to another case simply moving at any speed, as well as the interest that such a phenomenon would have in obtaining an important metric deformation.
An illustration of the difference between these two cases is proposed in Figure \ref{CylindreMob}b.

\indent \textit{Note:} In any other case where the cylinder moves at speed $v<c$, one could have applied a Lorentz transform to the static case in order to take into account the new relative speed of the cylinder. However, this transform produces divergent values for $v=c$. We cannot use the Lorentz transform, so it is imperative to solve the problem by convolution product of the $Source(ct,z,r)$ term with the D'Alembertian Green function. 

In order to simplify the calculations, we will place ourselves in the comoving frame of reference in respect to the cylinder of light, that is to say:
\begin{equation} \label{ComobSet}
(O,ct,z,z')\mapsto (O,ct,Z,\tilde{z}) \text{ such as: } \left\{ \begin{array}{l}
Z=z-ct\\
\tilde{z}=z'-z\\
\end{array} \right.
\end{equation}
We take the expression of the solution given by the expression (\ref{SolGreen0}), and we now introduce the source term corresponding to a cylinder of light moving at $c$:
\begin{widetext}
\begin{equation}\label{GreenMobInt2}
\phi_0(ct,z)= {\cal{A}} \chi \int^{\infty}_{-\infty} d\tilde{z} \int^{\infty}_{0} H(ct-\rho_0)H(\tilde{z}+Z+\rho_0)H(L-Z-\tilde{z}-\rho_0)H(R-r') \dfrac{r' dr'}{\rho_0}
\end{equation}
Where $\rho_0(r',\tilde{z})=\sqrt{\tilde{z}^2+r'^2}$. Which leads us, by the same reasoning as the one presented in part \ref{sec:CylImmob}, to: 
\begin{equation}
\begin{split}
& \phi_0(t,z,0)= {\cal{A}} \chi \int^{\infty}_{-\infty} d\tilde{z} H(\tilde{z}+ct) H(ct-\tilde{z}) H(L-Z) H(L-Z-2\tilde{z}) \times \\
& \times \left[\begin{array}{c}
\text{\small upper}\\
\text{\small integration}\\
\text{\small bound}\\
\end{array} \left\lbrace\begin{array}{l}
H((L-Z-\tilde{z})^2-c^2t^2) H(R^2+\tilde{z}^2-c^2t^2) H(c^2t^2 - (\tilde{z} +Z)^2) ~ ct \\
 + H(c^2t^2-(L-Z-\tilde{z})^2) H(R^2+\tilde{z}^2-(L-Z-\tilde{z})^2)~ (L-Z-\tilde{z})\\
  + H(c^2t^2-z^2-R^2) H((L-Z-\tilde{z})^2-z^2-R^2) H(R^2 + Z(-Z-2\tilde{z}))~ \sqrt{R^2+\tilde{z}^2}\\
  \end{array}\right. \right.\\
& \hspace{3.4 cm} \left. - \underbrace{ H(-Z)H(-Z-2\tilde{z})~ (-Z-\tilde{z}) - (H(Z)+H(-Z)H(Z+2\tilde{z}))~  |\tilde{z}|}_{\text{\small{lower integration bound}}} \right]
\end{split}\label{eq:hMobDev}
\end{equation}
\end{widetext}
To which equation follows a calculation step applied to a Python program having the same function of solution representation as the one presented part \ref{sec:CylImmob}.
%%%%%%%%%%%%%%%%%%%%%%%%%%%%%%%%%%%%%%%%%%%%%%%%%%%%%%%%%%%%%%%%
\section{Results and analysis}
\subsection{Study at constant intensity and energy}
\begin{figure}
\begin{center}
\begin{minipage}{\linewidth}
\includegraphics[width=\linewidth]{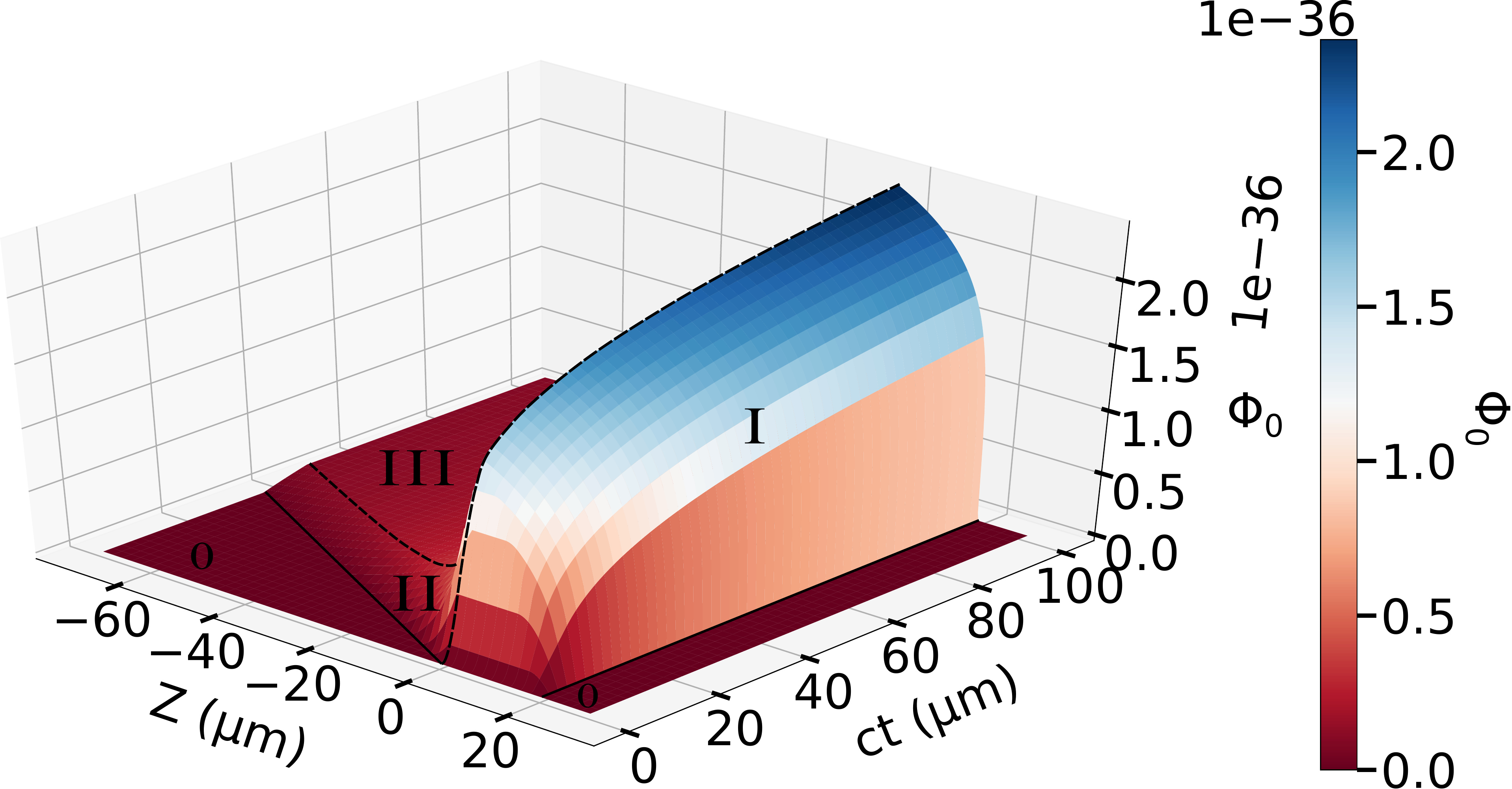}
\end{minipage}
\begin{minipage}{\linewidth}
\includegraphics[width=\linewidth]{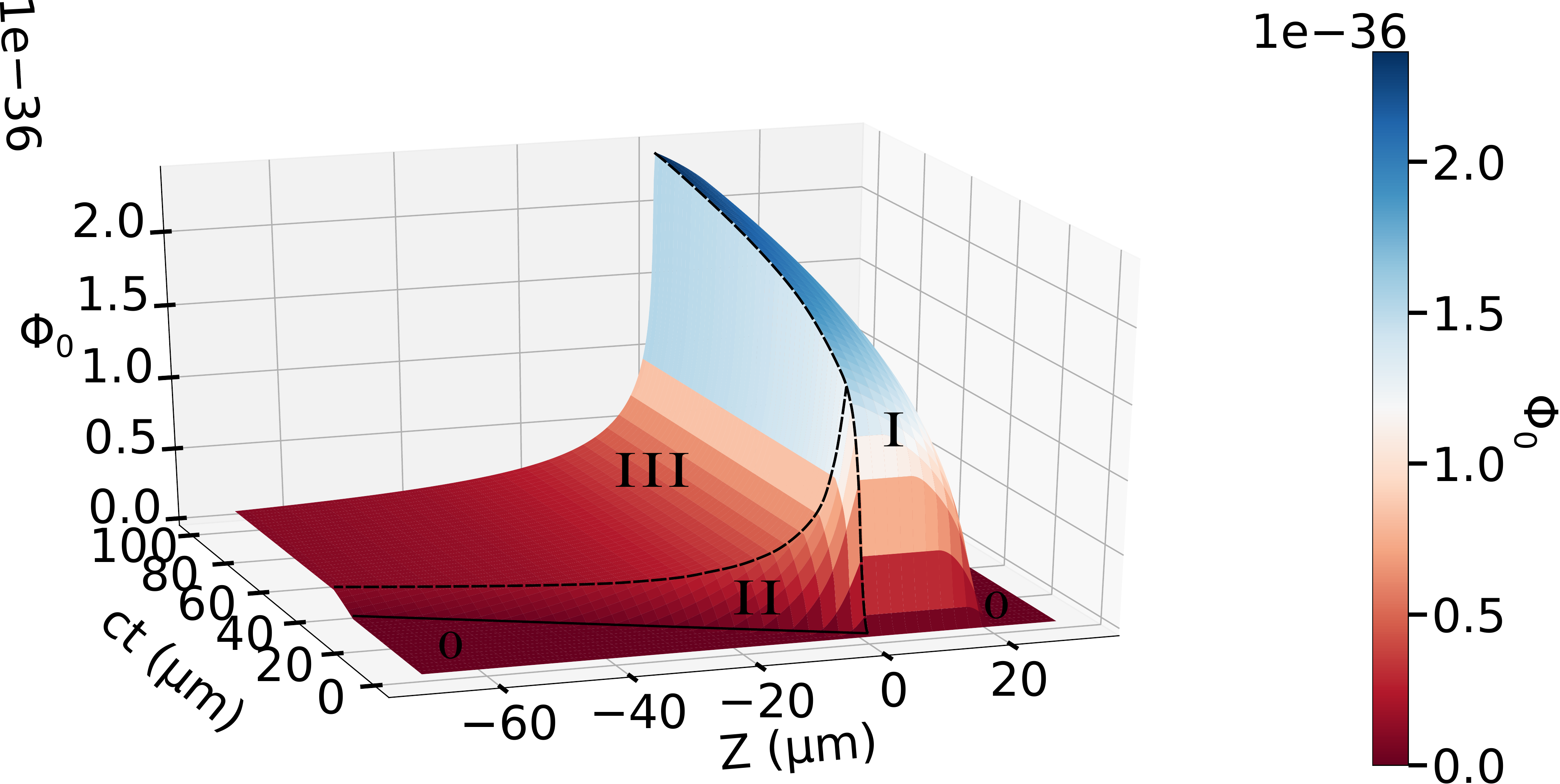}
\end{minipage}
\caption{3D representation according to two angles {\large (} "from the front" and "from the back"{\large )} of the solution on the comoving plane $(O,Z,ct)$ for a laser intensity $I=10^{22}~\rm W/cm^2$ and dimensions of the light cylinder $L=20~\rm~\mu m,~ R=~5\rm~\mu m$. Zone I/ the growth regime inside the source, zone II/ the transient regime outside the source, zone III/ the quasi-static regime outside the source and zone 0/ unperturbed spacetime.\label{Ex3DMob}}
\end{center}
\end{figure}

\begin{figure}[h!]
\begin{center}
\begin{minipage}{\linewidth}
\includegraphics[width=\linewidth]{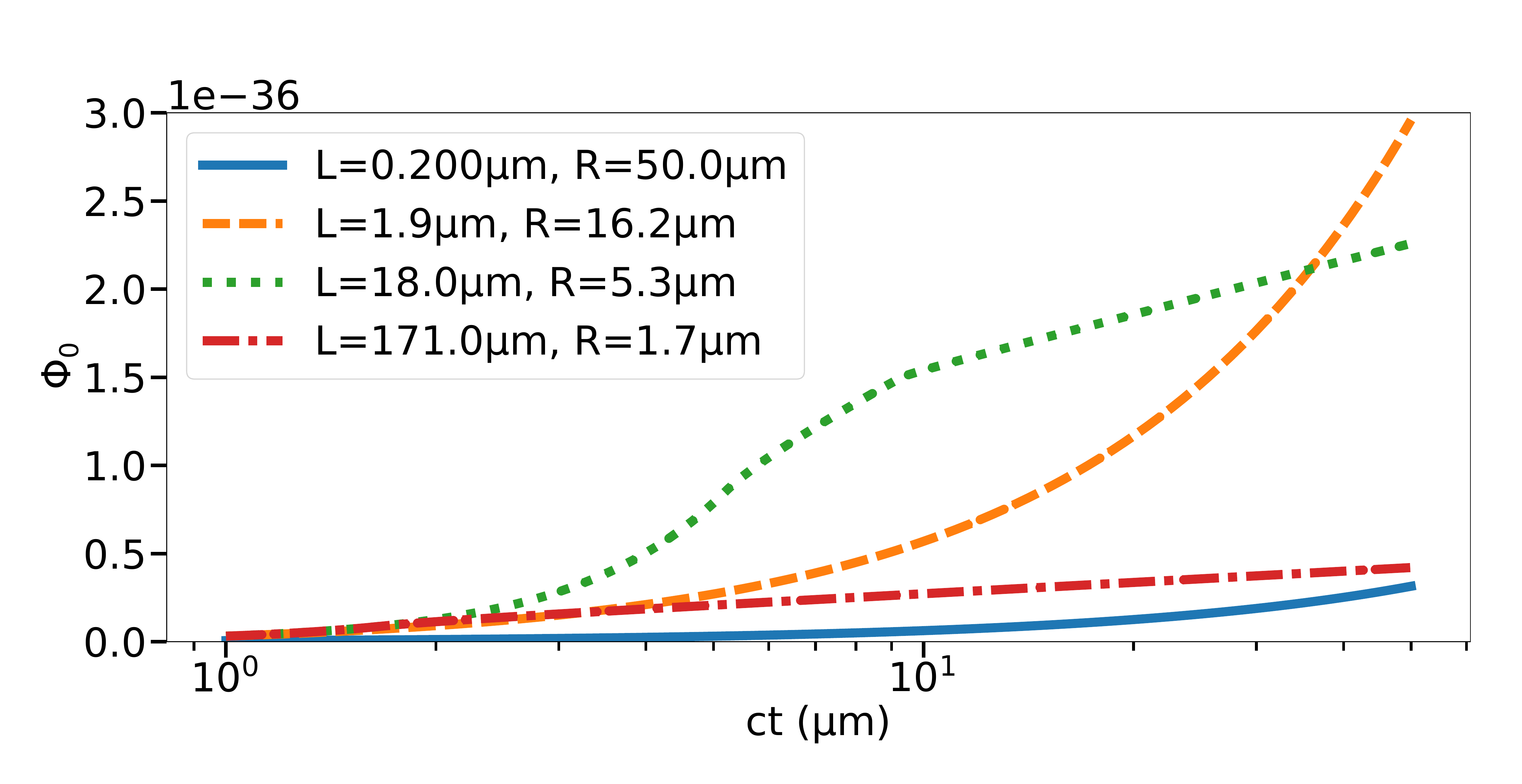}
\end{minipage}
\begin{minipage}{\linewidth}
\includegraphics[width=\linewidth]{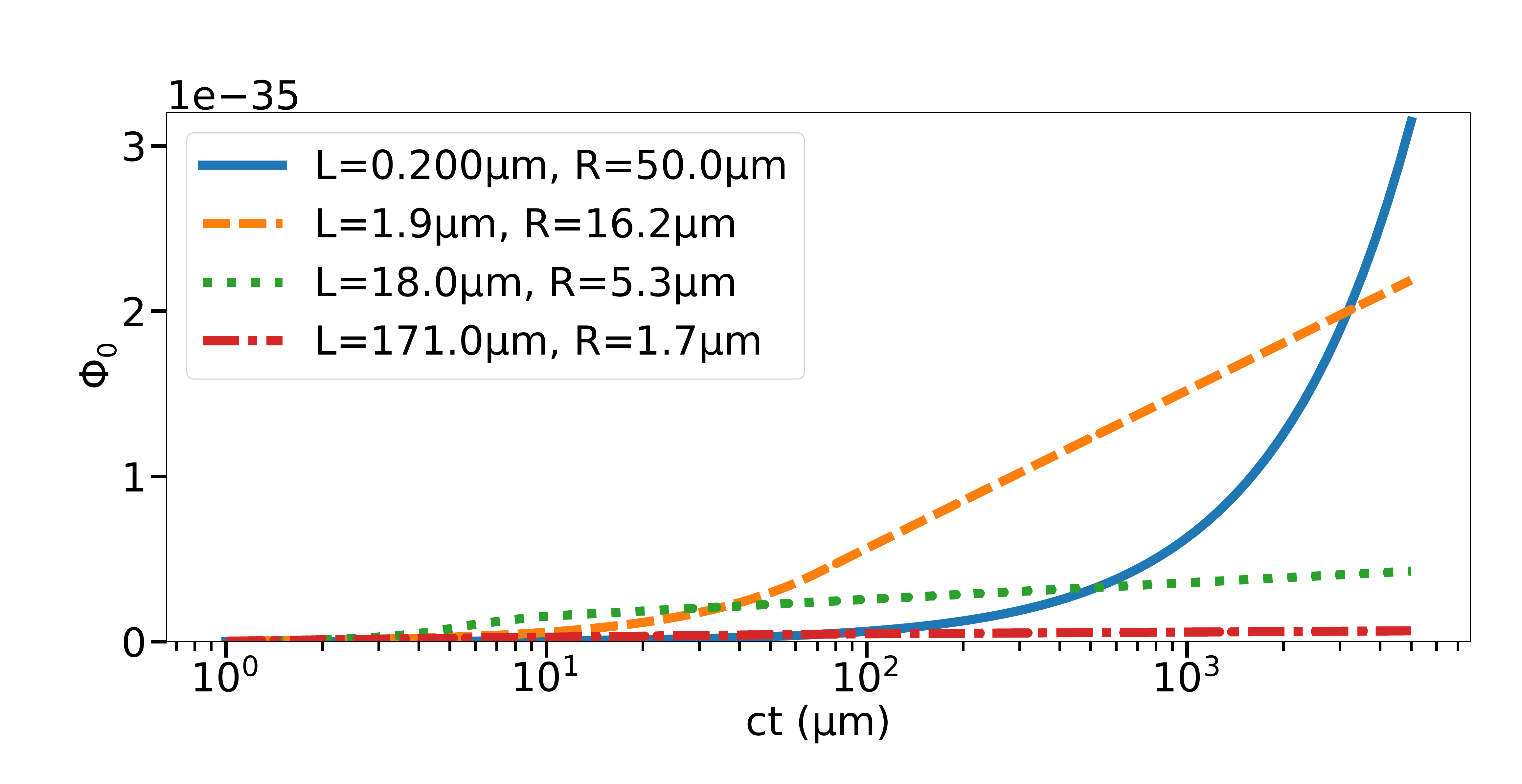}
\end{minipage}
\caption{Detail of the growth of the maximum of metric perturbation until the time such as a) $ct=50~\rm \mu m$ b) $ct=5~\rm mm$. This perturbation is generated by a cylinder of intensity $I=2 \times 10^{22}~\rm W/cm^2$ and energy $E=525~\rm J$ as a function of time for cylinders with various aspect ratios. \label{xsliceMob}}
\end{center}
\end{figure}

\begin{figure}
\begin{center}
\includegraphics[width=\linewidth]{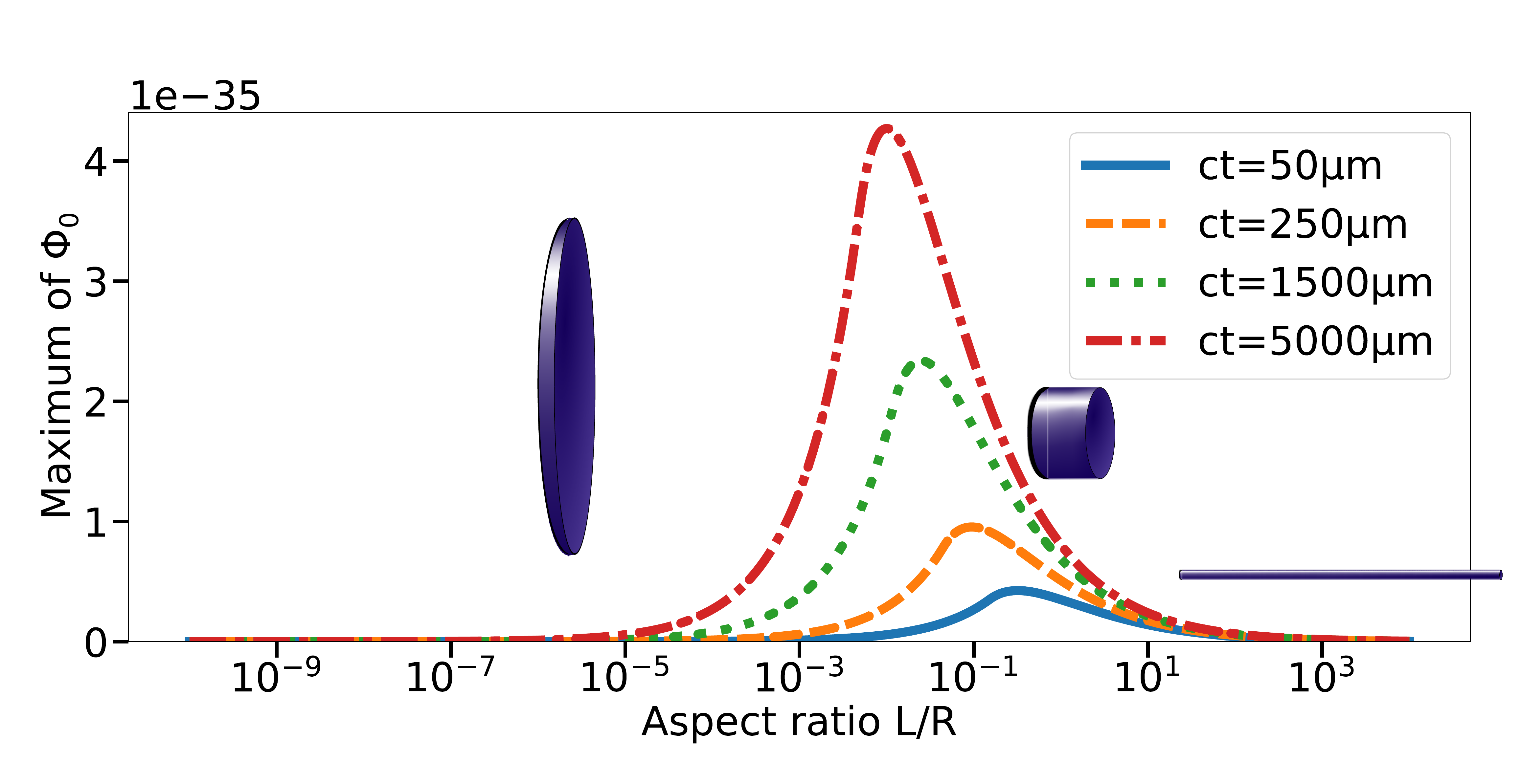}
\caption{Maximum amplitude at fixed time of the metric perturbation generated by a cylindrical light pulse of intensity $I=2 \times 10^{22} \rm ~W/cm^2$ and energy $E=525 \rm~J$ according to the aspect ratio L/R of said cylinder.\label{maxMob}}
\end{center}
\end{figure}

In order to model again what could be obtained with a current ultra high power laser \citep{danson_petawatt_2019} , we take again as characteristics $I={\cal{A}} \times c= 10^{22}~ \rm W/cm^2$, $E=525~\rm J$, which still correspond to the dimensions $R=~5~\rm \mu m$ and $L=~20\rm~\mu m$. The Python program then gives us the visualization shown in Figure \ref{Ex3DMob}.
The metric perturbation generated by a cylinder of light moving at speed $c$ has a characteristic "wave" profile, instead of the symmetrical profile we had in the static case. Indeed, the fact that we place ourselves in the comoving frame implies that, by causality, no perturbation can be present "in front of" the source, i.e. for $Z>L$. The maximum is located at the comoving coordinate $Z=0$ in a rather logical way. Indeed, any perturbation generated in the source at a comoving coordinate $Z_1$ cannot be observed at a later time at a point of comoving coordinate $Z_2>Z_1$, because the opposite would imply that a perturbation has moved faster than the speed of light. Still, we keep a $1/Z$ shaped potential in the "trail" of this wave (zone III). Among the differences that can be observed between the static case and this new mobile case, is the growth of the disturbance in the source, which instead of reaching a stationary regime after a certain time, seems to continue to grow in zone I. Indeed, a logarithmic growth now seems to appear after the first growth phase that we could already observe in the static case. This growth can be easily isolated in the different terms of contribution to the perturbation.
%%%%%%%%%%%%%
Thus for $Z=0$, at long times such that $ct \gg R$ and $R \gg L$, one can isolate as the only contribution to the perturbation the following terms:
\begin{align}
\phi_0 =  \dfrac{\chi {\cal{A}}}{2} &\left(\dfrac{R^2}{2} + R^2 \left[\text{arcsh}\left( \dfrac{L}{2R}-\dfrac{R}{2L} \right) \right.\right. \\
&\left. \left. + \text{arcsh}\left(\sqrt{\dfrac{c^2t^2}{R^2}-1} \right) \right] +ct\left(ct-\sqrt{c^2t^2-R^2} \right) \right) \notag
\end{align}
Which nets us through a Taylor expansion:
\begin{equation}
\phi_0 \approx \dfrac{\chi {\cal{A}}R^2}{2} \left[ 1 + \text{ln}\left( \dfrac{4Lct}{R^2} \right) \right] \label{DLln}
\end{equation}
In order to supplement this expansion carried out for a limited domain of space-time, we will again be interested in the impact of the shape of the cylinder on the amplitude at a given time of this deformation. The goal is finally to find the optimal configuration to generate the largest possible deformation.
For this, we will study the profile of the maximum of this deformation, located at $Z=0$ for several aspect ratios, as shown in Figure \ref{xsliceMob}.
We observe that each maximum of metric deformation seems to evolve in time according to two modes of growth: in a first period of time, the perturbation sees an acceleration of its growth (in the broad sense, this growth is at least linear), followed by a second period, with longer times, where the growth becomes logarithmic. The duration of this first period of time is different for each aspect ratio, and appears to occur later and later as the aspect ratio decreases. Thus, the cylinder of light with the second largest aspect ratio ($L=18\rm ~\mu m$, $R=5.3 \rm~\mu m$), initially generates the highest maximum. But the slowing of its growth at longer times means that at $ct=50 \rm~\mu m$ the largest deformation maximum is now held by another cylinder of light ($L=1.9 \rm ~\mu m$, $R=16.2 \rm~\mu m$). By the same logic, the cylinder of light with the smallest aspect ratio ends up catching up with the previous cylinder in terms of maximum generated perturbation. We must therefore expect to see the optimal aspect ratio of the cylinder of light, i.e. the one which gives the greatest metric perturbation on the axis, move towards the small aspect ratios as time increases. This hypothesis is confirmed by plotting the maximum perturbation at a given time as a function of the aspect ratio in Figure \ref{maxMob}.
We can observe on this occasion that the maximum of deformation increases with time in a way that cannot only be explained by the logarithmic growth highlighted earlier. It seems that taking a cylinder of light of larger radius, and therefore of smaller length at constant intensity and energy, favors the generation of a larger metric perturbation. This can be explained by the fact that such a "light disk" has at long times more source points, and thus emitting points, close to the observation point in $Z=0$ at the time of the observation.

Thus, in an ideal setting where one could make a light pulse propagate as a cylinder of energy density over an indefinite time, the intensity plays an important role since the amplitude grows linearly with it. Nevertheless, another way to maximize the metric deformation at constant intensity could also be to let the perturbation grow by letting the light propagate as long as possible, under the explicit condition of adjusting the aspect ratio of the cylinder to the distance traveled to have the best efficiency. The amplitude of this deformation will be limited by the length on which we can experimentally focus the beam.

This deformation's front has a peculiar wave shape, which ensures a strong instantaneous variation of the metric when the metric deformation would reach the observer. Now the variations of the metric are responsible for some of the effects observable in gravity potentials, as the deflection of light. The deformation's profile has therefore interesting properties for the measurement of a metric perturbation generated in the laboratory.
%%%%%%%%%%%%%%%%%%%%%%%%%%%%%%%
\subsection{Study at constant power}
Most laser facilities propose lasers with a fixed maximum power. We must therefore observe the dependence of the metric deformation on power.\\
We thus study the metric deformation generated at fixed power $P= 1 \, \rm PW$, as it is again a realistic power considering current high-power lasers \cite{danson_petawatt_2019}. 

In this case, at constant power, $L$ does not change the intensity of the beam, since $P=\pi R^2 I$. We will however represent the deformation at different L, to check if the change of aspect ratio has an influence on the observed deformation. Equation \eqref{DLln} seems to show that at long time for $R\gg L$ the solution could be proportional to $IR^2$ and thus to the power $P$. \\
If this proportionality is proven in a more global framework, we would end up with a parameter to be adjusted, the power, not requiring the extreme focusing of the laser beam and thus avoiding the complications arising with very high intensity light beams.

\begin{figure}[h!]
\begin{center}
\includegraphics[width=\linewidth]{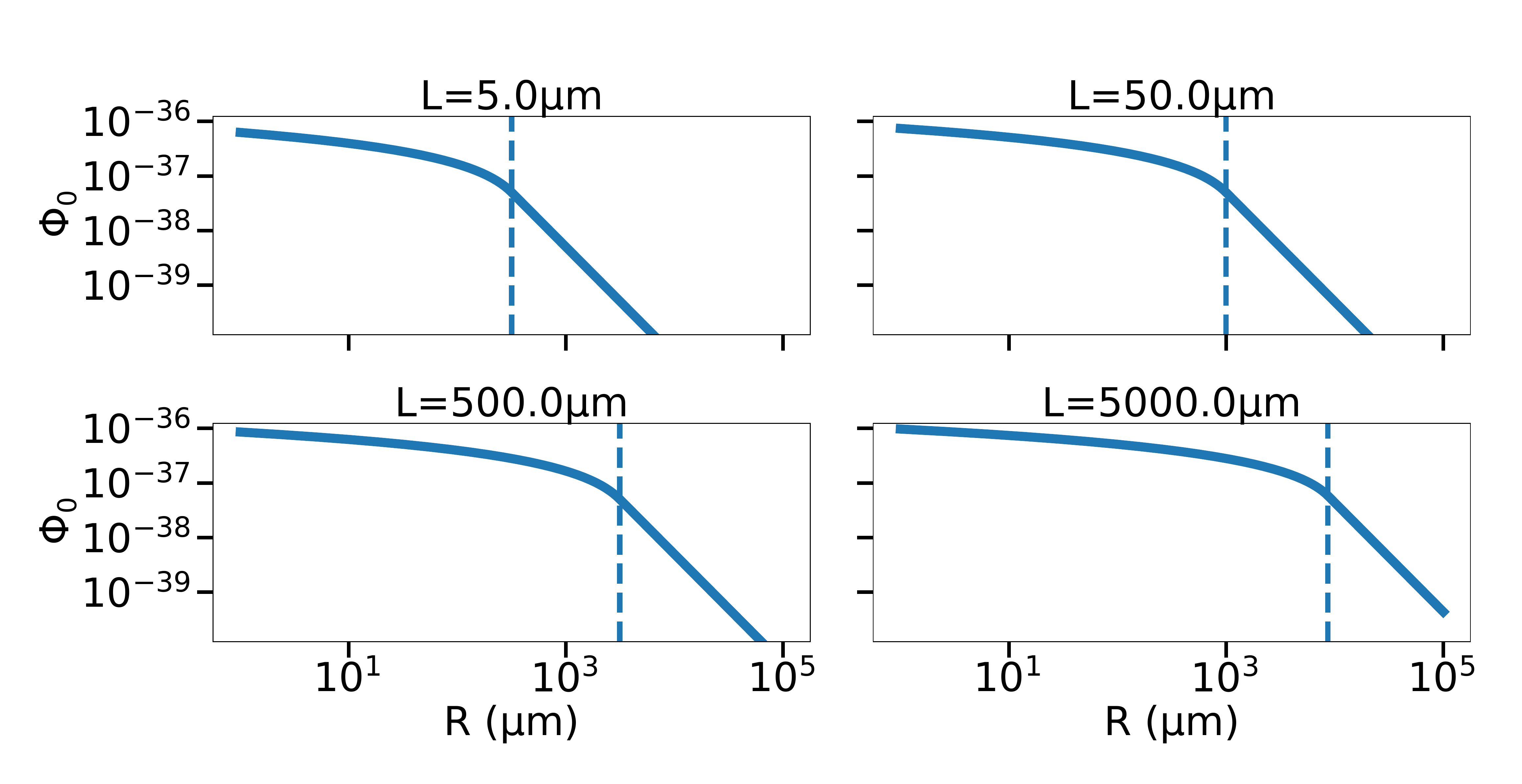}
\caption{Maximum of amplitude at time such that $ct=1 \rm~cm$ of the metric perturbation of a cylindrical light pulse of fixed length $L$ (i.e. at fixed time) {\large (}$L=5 \rm~\mu m$, $L=50 \rm \mu m$, $L=500 \rm~\mu m$, $L=5000 \rm~\mu m${\large )} and of constant power $P=1 \rm~PW$ as a function of the radius $R$ of the cylinder.\label{maxPcte}}
\end{center}
\end{figure}

At fixed time the maximum of metric perturbation seems to decrease when the radius increases, as presented in Figure \ref{maxPcte}. This result may seem at first sight contradictory to the result of Equation \eqref{DLln}, but two differences must be taken into account here. The plots are done at constant time, as the radius R of the cylinder is what varies. Now, the equation includes a term $\text{ln}\left( 4L/R^2 \right)$ which decreases as $R$ increases. But there is also a radius $R$ from which we no longer have $ct \gg R$ and where the Taylor expansion is no longer valid.
At large R's, we can relate this decrease to the relation $I=P/(\pi R^2)$. Indeed, $R$ being larger than $ct$, the deformation generated by the edges of the cylinder could not yet reach the center of the cylinder. So we only decrease the intensity of the beam without really increasing the volume of contributions.
However, in a situation closer to physical reality, a light beam does not remain cylindrical \textit{ad eternam}. It can only be considered cylindrical over twice its Rayleigh length, that is, twice the distance it takes for a focused beam to diffract and see its radius multiplied by $\sqrt{2}$. This distance depends of course on how focused the beam is, as well as its wavelength, according to the designated formula:
\begin{equation}
z_r=\dfrac{\pi R^2}{\lambda}
\end{equation}
Where $z_r$ is the Rayleigh length, $R$ the minimum radius of the beam, and $\lambda$ the focused light's wavelength.
This quantity thus introduces an upper limit on the propagation time of the cylinder of light, which is $t_r=2 z_r/c$. There is therefore a direct link between the maximum propagation time and the radius $R$ of the cylinder. If we consider that we want the largest possible perturbation of the space-time metric, we can then consider that any cylinder of light that we study generates a metric perturbation over the longest possible time, that is $t_r$. We can then re-study the results at constant power, as a function of $R$, such that the cylinder of light has generated a metric deformation over $2z_r$.
\begin{figure}
\begin{center}
\includegraphics[width=\linewidth]{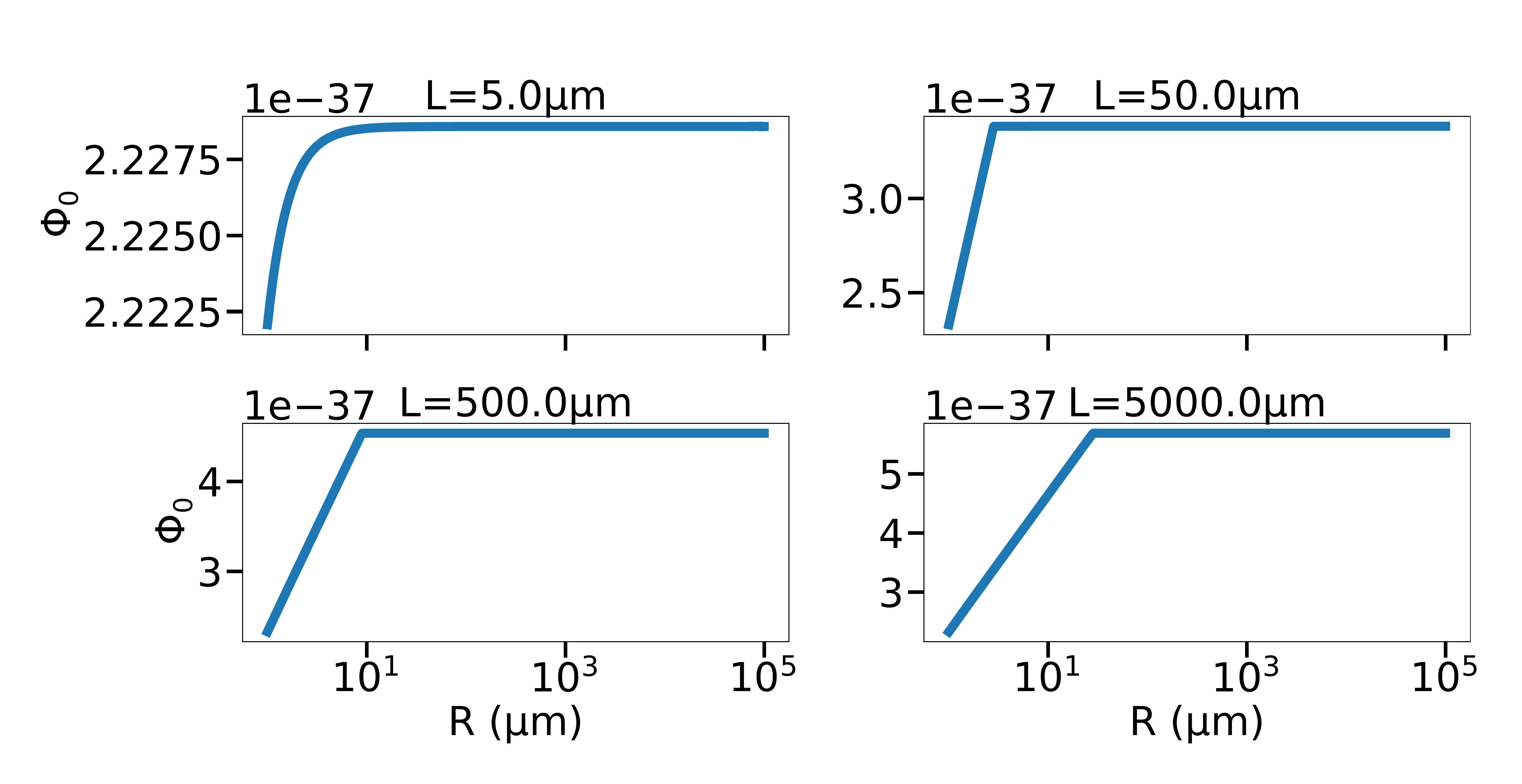}
\caption{Maximum of amplitude at set $L$ {\large (}$L=5 \rm~\mu m$, $L=50 \rm \mu m$, $L=500 \rm~\mu m$, $L=5000 \rm~\mu m${\large )} of the metric deformation at constant power $P= 1~\rm PW$ as a function of the cylinder radius $R$ and at time such that $ct=2~z_R$\label{maxPcteRt}}
\end{center}
\end{figure}
This new calculation which takes more into account the experimental situation, and presented in Figure \ref{maxPcteRt}, shows that the increase of the cylinder radius is far from being detrimental, contrary to what was previously thought. As long as the radius $R$ is sufficiently large compared to the length $L$ of the cylinder, the metric deformation generated by a cylinder of light is indeed constant at constant power.

Taking Equation \eqref{DLln} and introducing $t_r$, we then obtain:
\begin{equation}\label{eq:hZr}
    \phi_0 \approx \dfrac{\chi P}{2 \pi c} \left[ 1 + \text{ln}\left( \dfrac{8\pi L}{\lambda} \right) \right]
\end{equation} 
This last expression is, at constant power $P$, independent of $R$ and logarithmically dependent on the ratio $L/\lambda$. These results confirm, in particular, that the determining physical quantity for the metric deformation is the power $P$ of our light beam.
In practice, this very important result allows us to get rid of the need for a very focused light beam, and thus of the quantum electrodynamics effects appearing at ultra high intensities. It also allows us to have experiment sizes that can be calibrated to the detection method used.
%%%%%%%%%%%%%%%%%%%%%%%%%%%%%%%%%%%%%%%%%%%%%%%%%%%%%%%%%%%%
\section{Resolution for an oscillatory part}
We then shift our focus to the oscillatory part of the source energy-stress tensor presented in equation \eqref{eq:hSource}, which can be reduced to the oscillatory solution $\phi_{osc}$ of equation \eqref{eq:hSourceOsc}.

\begin{equation} \label{eq:hSourceOsc}
    -\square \phi_{2k} = \chi \epsilon_0 E_0^2 \cos(2k(z-ct)) 
\end{equation}

A physical interpretation of such an equation can be seen as the transverse perturbation of the metric generated by the interaction of an electromagnetic pulse with a static electromagnetic field such that the extent of the electromagnetic field is large before the spatial and temporal extension of the pulse. It is the result of a Gertsenshtein \cite{gertsenshtein_wave_1962} effect where we take a light pulse of small finite dimension.

We consider once again a cylinder going at the speed of light $c$ but which stress moment is now oscillatory. We can now write the source term in equation \eqref{eq:hSourceCte} as:

\begin{equation} \label{eq:SrcOsc}
\begin{split}
    S & (ct,z,r,\theta)= {\cal{A}} \cos(k(z-ct)) \times\\
    & \times H(ct)H(R-r)H(r) H(z-ct)H(L-(z-ct))
\end{split}
\end{equation}

Applying the methods of calculations we have seen in the previous part, it is possible to determine once again the analytic expression of $\phi_{osc}$. The results for the integration on the Heaviside functions is the same as is described equation \eqref{eq:hMobDev}, only now the integration on the functions changes and is now dependant on sinus functions. Setting ourselves in the comoving set of coordinates described in expression \eqref{ComobSet}, we thus get after the first integration the equivalent of, for a cylinder of oscillatory moment, equation \eqref{eq:hMobDev}:

\begin{widetext}
\begin{equation}
\begin{split}
& \phi_{k}(t,z,0)= \dfrac{{\cal{A}} \chi}{k} \int^{\infty}_{-\infty} d\tilde{z} H(\tilde{z}+ct) H(ct-\tilde{z}) H(L-Z) H(L-Z-\tilde{z}) \times \\
& \times \left[\begin{array}{c}
\text{\small upper}\\
\text{\small integration}\\
\text{\small bound}\\
\end{array} \left\lbrace\begin{array}{l}
H((L-Z-\tilde{z})^2-c^2t^2) H(R^2+\tilde{z}^2-c^2t^2) H(c^2t^2 - (\tilde{z} +Z)^2)~ \sin(k(Z+\tilde{z}+ct)) \\
 +  H(c^2t^2-(L-Z-\tilde{z})^2) H(R^2+\tilde{z}^2-(L-Z-\tilde{z})^2)~ \sin(k(Z+\tilde{z}+L-Z-\tilde{z}))\\
  + H(c^2t^2-z^2-R^2) H((L-Z-\tilde{z})^2-z^2-R^2) H(R^2 + Z(-Z-2\tilde{z}))~ \sin(k(Z+\tilde{z}+\sqrt{R^2+\tilde{z}^2}))\\
  \end{array}\right. \right.\\
& \hspace{1.8 cm} \left. - \underbrace{ H(-Z)H(-Z-2\tilde{z})~ \sin(k(Z+\tilde{z}-Z-\tilde{z})) - (H(Z)+H(-Z)H(Z+2\tilde{z}))~  \sin(k(Z+\tilde{z}+|\tilde{z}|))}_{\text{\small{lower integration bound}}} \right]
\end{split}\label{eq:hMobDev}
\end{equation}
\end{widetext}

\begin{figure}
\begin{center}
\includegraphics[width=\linewidth]{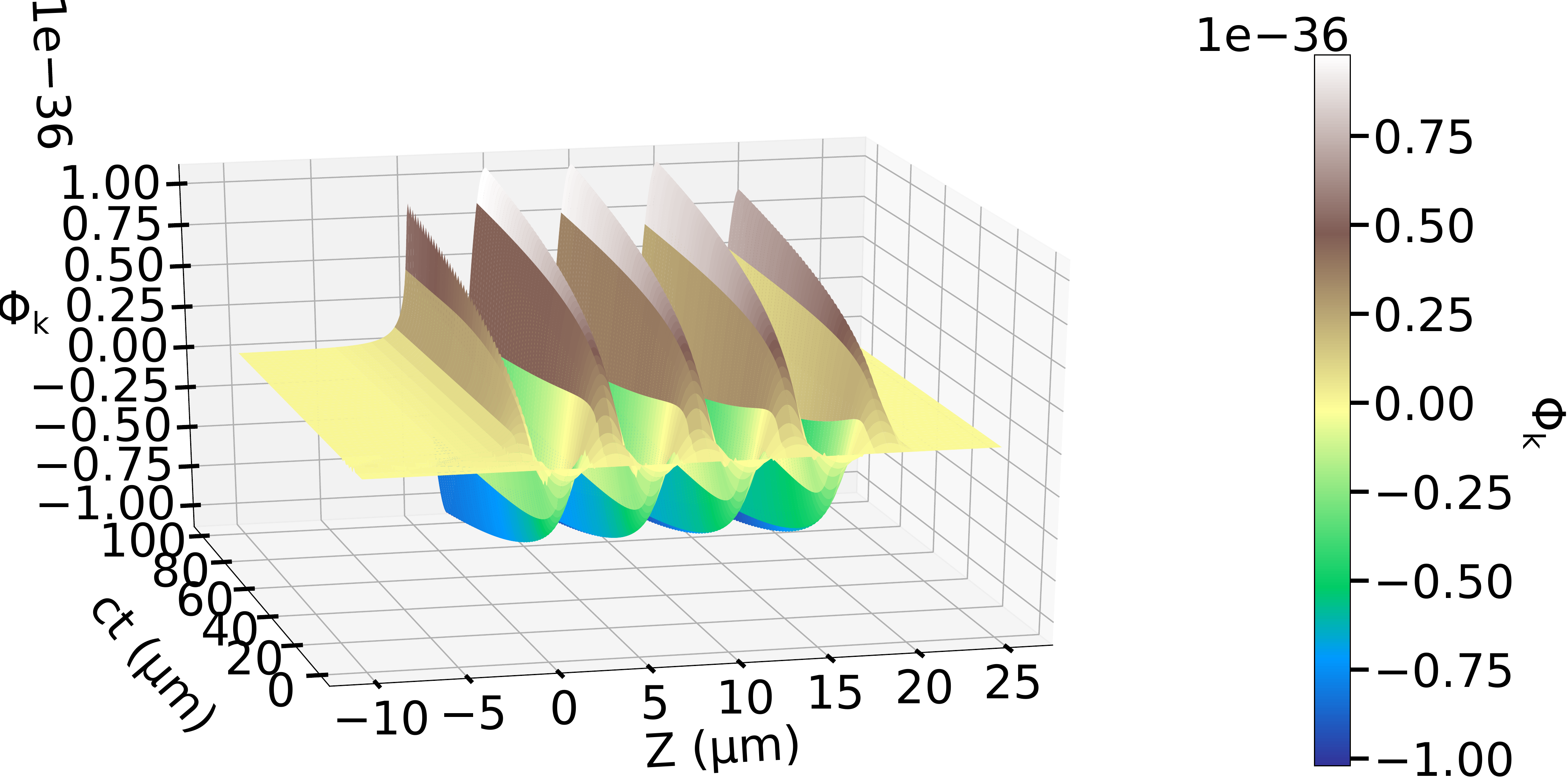}
\caption{3D representation on the comoving plane of the solution for an oscillating source term for a laser intensity $I=10^{22}~\rm W/cm^2$ and dimensions of the light cylinder $L=20~\rm~\mu m,~ R=~5\rm~\mu m$.\label{3DOscill}}
\end{center}
\end{figure}

Then, applying the same algorithmic method as usual we obtain the profile Figure \ref{3DOscill} of the metric deformation generated by an electromagnetic pulse with characteristics $I={\cal{A}} \times c= 10^{22}~ \rm W/cm^2$, $E=525~\rm J$, and $R=~5~\rm \mu m$ the radius and $L=~20\rm~\mu m$ the length of cylinder. We take the wavelength of the source oscillation as $\lambda=2\pi /k = 5 \, \rm \mu m$.

We observe a reproduction of the electromagnetic oscillation as a synchronous oscillation of the metric perturbation. This differentiates the oscillatory case from the constant source case as we don't get the same wave-shaped envelope for the metric deformation in the $Z$ direction, even though we get the same logarithmic growth profile in time for both cases. Thus, instead of having the maximum of deformation in $Z=0$, we can study the overall amplitude of metric perturbation as an estimate of the maximum of deformation produced.

\begin{figure}
\begin{center}
\includegraphics[width=\linewidth]{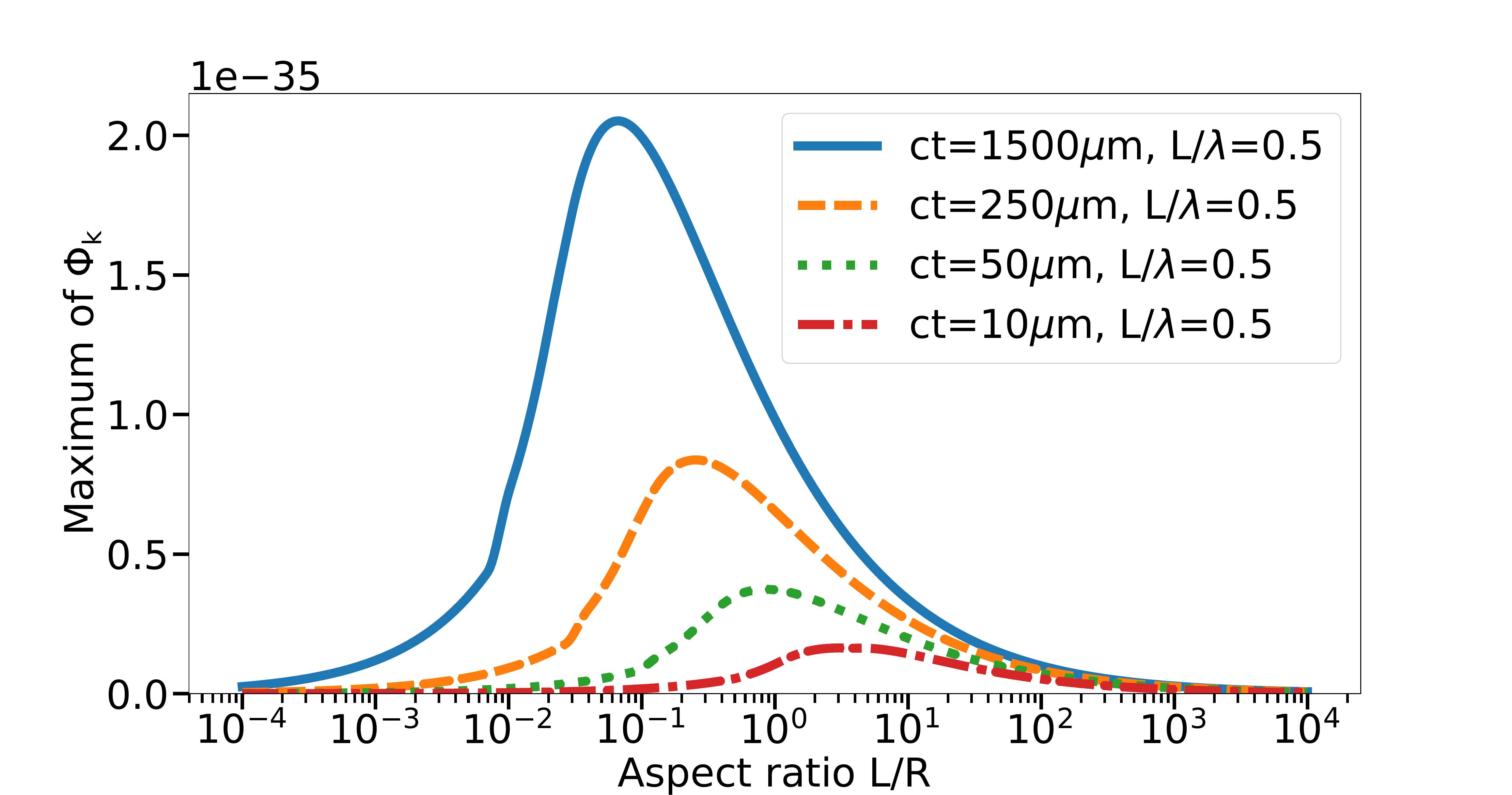}
\caption{Maximum amplitude at fixed number of optical cycles $L/\lambda=0.5$, and for varying times, of the metric perturbation generated by a cylindrical light pulse of intensity $I=2 \times 10^{22} \rm ~W/cm^2$ and energy $E=525 \rm~J$ according to the aspect ratio L/R of said cylinder.\label{maxOscill_ct}}
\end{center}
\end{figure}

Let us then study this amplitude for different aspect ratios, times, and wavelengths, at set intensity and energy, as we did in the previous part.\\
First considering the problem at set wavelength such that it gives the smallest possible number of optical cycles in the cylinder $L/\lambda=0.5$, we study the influence of time and aspect ratios on the amplitude of the metric perturbation, as it was before studied for the constant part of the source in Figure \ref{maxMob}.
What we thus observe in Figure \ref{maxOscill_ct} is fairly similar to the previously studied constant source case as we get the same behavior, both in time and in aspect ratio, as in Figure \ref{maxMob}. As the interaction time grows, so does the maximum of amplitude, but the position of that maximum shifts towards lower aspect ratios for which the cylinder radius $R$ gets larger, and the length $L$ smaller. We can thus safely assume that this oscillatory case behaves mostly the same way as its constant source counterpart, for considerations that do not depend on the wavelength.

\begin{figure}
\begin{center}
\includegraphics[width=\linewidth]{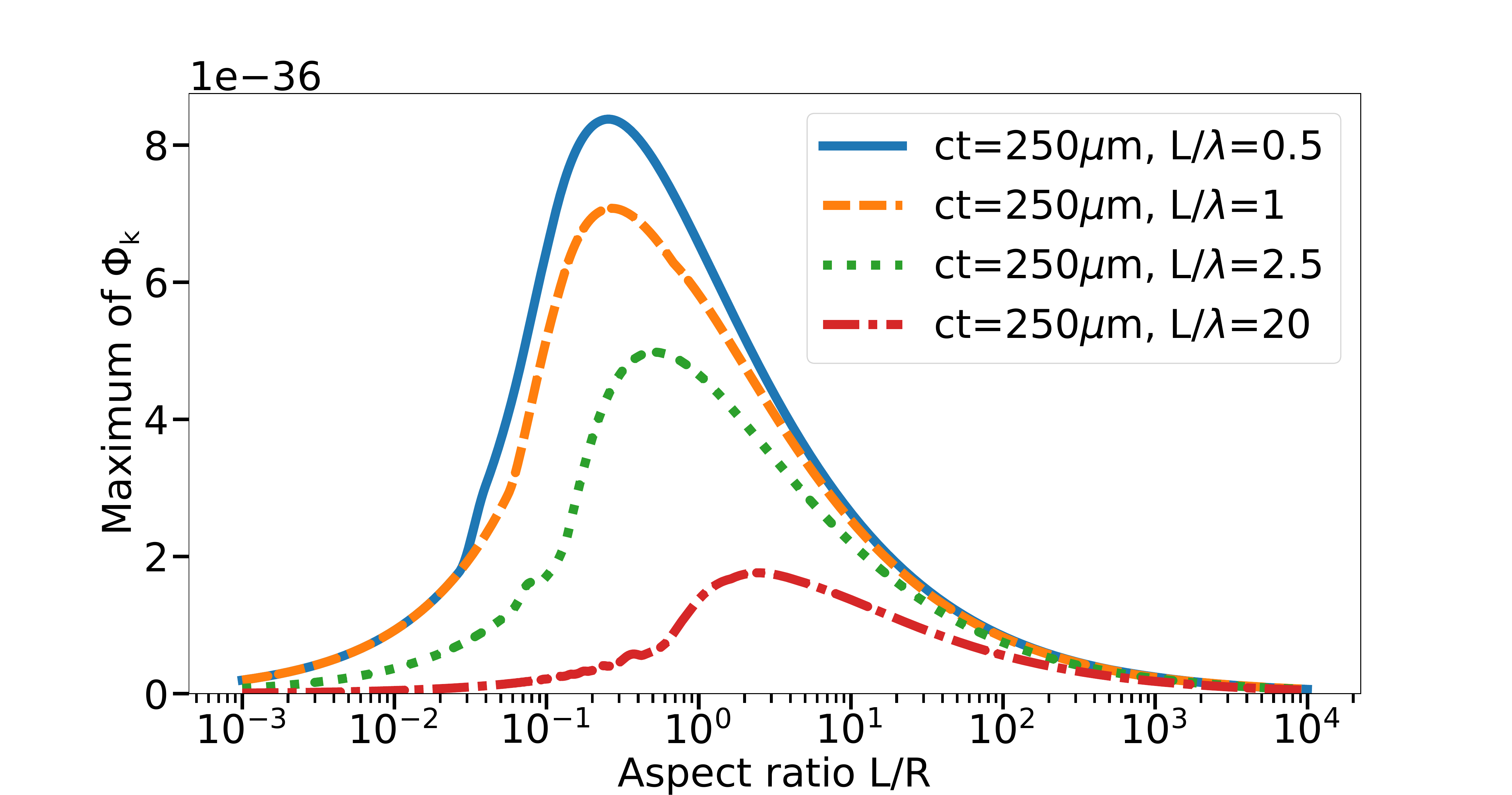}
\caption{Maximum amplitude at fixed time such as $ct=250 \rm \, \mu m$ and a varying number of optical cycles $L/\lambda=0.5$ of the metric perturbation generated by a cylindrical light pulse of intensity $I=2 \times 10^{22} \rm ~W/cm^2$ and energy $E=525 \rm~J$ according to the aspect ratio L/R of said cylinder.\label{maxOscill_k}}
\end{center}
\end{figure}

Quite naturally, we then study the influence of source wavelength on this same amplitude of the metric perturbation for a set time. As a dimensionless number linked to both wavelength and a physical aspect of electromagnetic oscillation, we will take instead of the wavelength the number of optical cycles $L/\lambda$, which must be integer or half-integer in order for the corresponding electromagnetic oscillation to satisfy the boundary conditions (i.e be zero) at the edges in the $Z$ direction of the cylinder. The results of such study are presented Figure \ref{maxOscill_k}.

As the number of optical cycles increases, the amplitude describes an opposite phenomenon to the one observed when the time increases: the maximum of amplitude gets smaller as $L/\lambda$ grows and the position of this maximum shifts towards higher aspect ratios.

It seems consequently that for set energy and intensity, we get the highest possible metric deformation for long times and the largest possible wavelength, which corresponds to an electromagnetic soliton.

\begin{figure}
\begin{center}
\includegraphics[width=\linewidth]{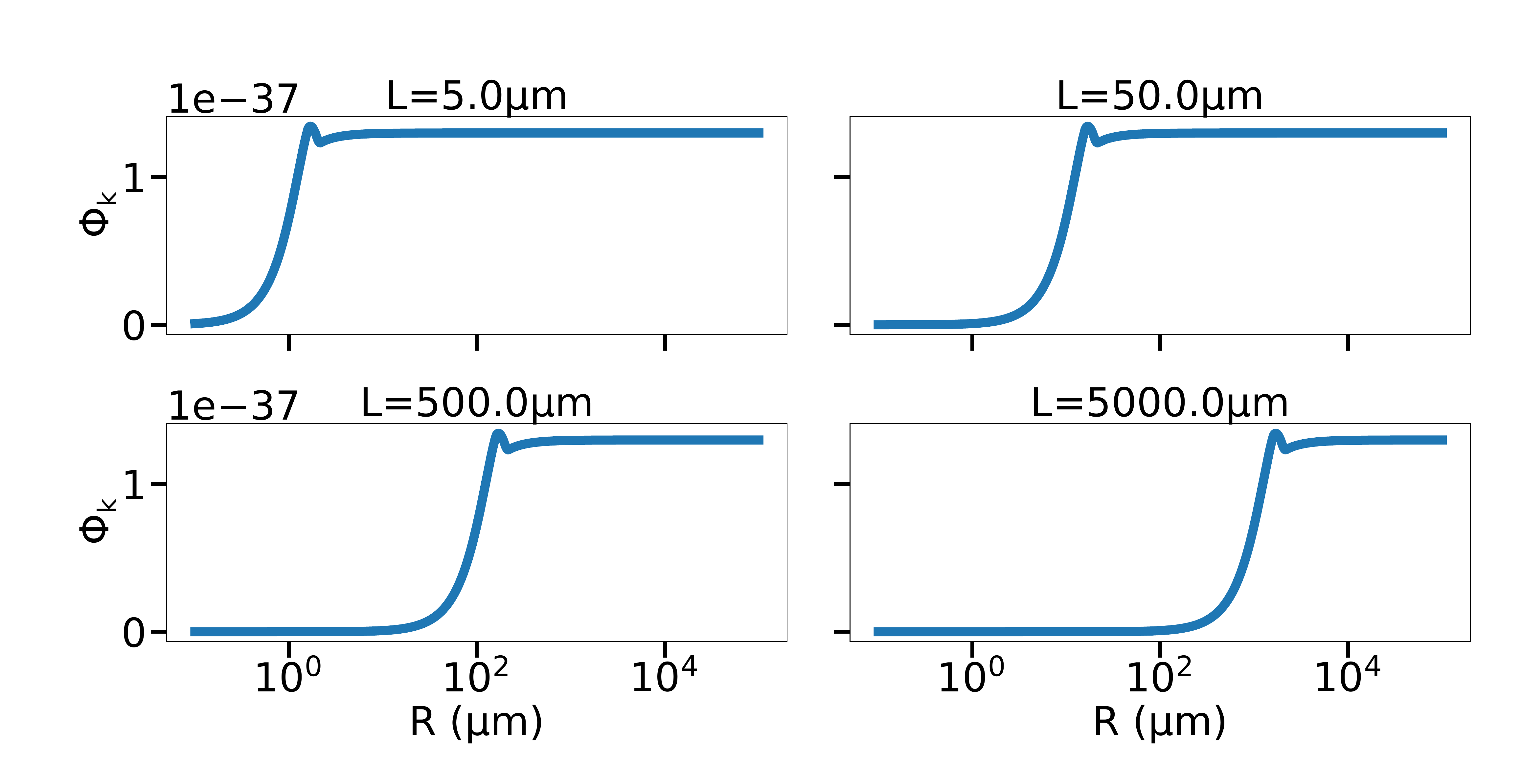}
\caption{Maximum of amplitude at set $L$ {\large (}$L=5 \rm~\mu m$, $L=50 \rm \mu m$, $L=500 \rm~\mu m$, $L=5000 \rm~\mu m${\large )} of the metric deformation at constant power $P= 1~\rm PW$ and number of optical cycles $L/\lambda =0.5$ as a function of the cylinder radius $R$ and at time such that $ct=2~z_R$.\label{maxPcteRtOscill}}
\end{center}
\end{figure}

As with the constant term, we now study what happens when we limit the time of propagation of this electromagnetic cylinder with diffraction, effectively reducing the domain of existence of the source cylinder to two Rayleigh lengths $2z_R$.
At set power $P=1 \rm \, PW$ and number of optical cycles $L/\lambda=0.5$, we unsurprisingly observe Figure \ref{maxPcteRtOscill} the same behavior as in the previous part. For $R>L$, the metric perturbation depends directly on the electromagnetic pulse power. It also does not seem to be dependent on neither the source wavelength nor the cylinder length $L$ as long as their ratio, the number of optical cycles $L/\lambda$, stays the same. This was already hinted by the Taylor expansion presented in the previous part equation \eqref{eq:hZr}, where the amplitude for the metric perturbation was only dependent on the power $P$ and the number of optical cycles $L/\lambda$, and it seems this consideration still holds for the oscillatory part.

\begin{figure}
\begin{center}
\includegraphics[width=\linewidth]{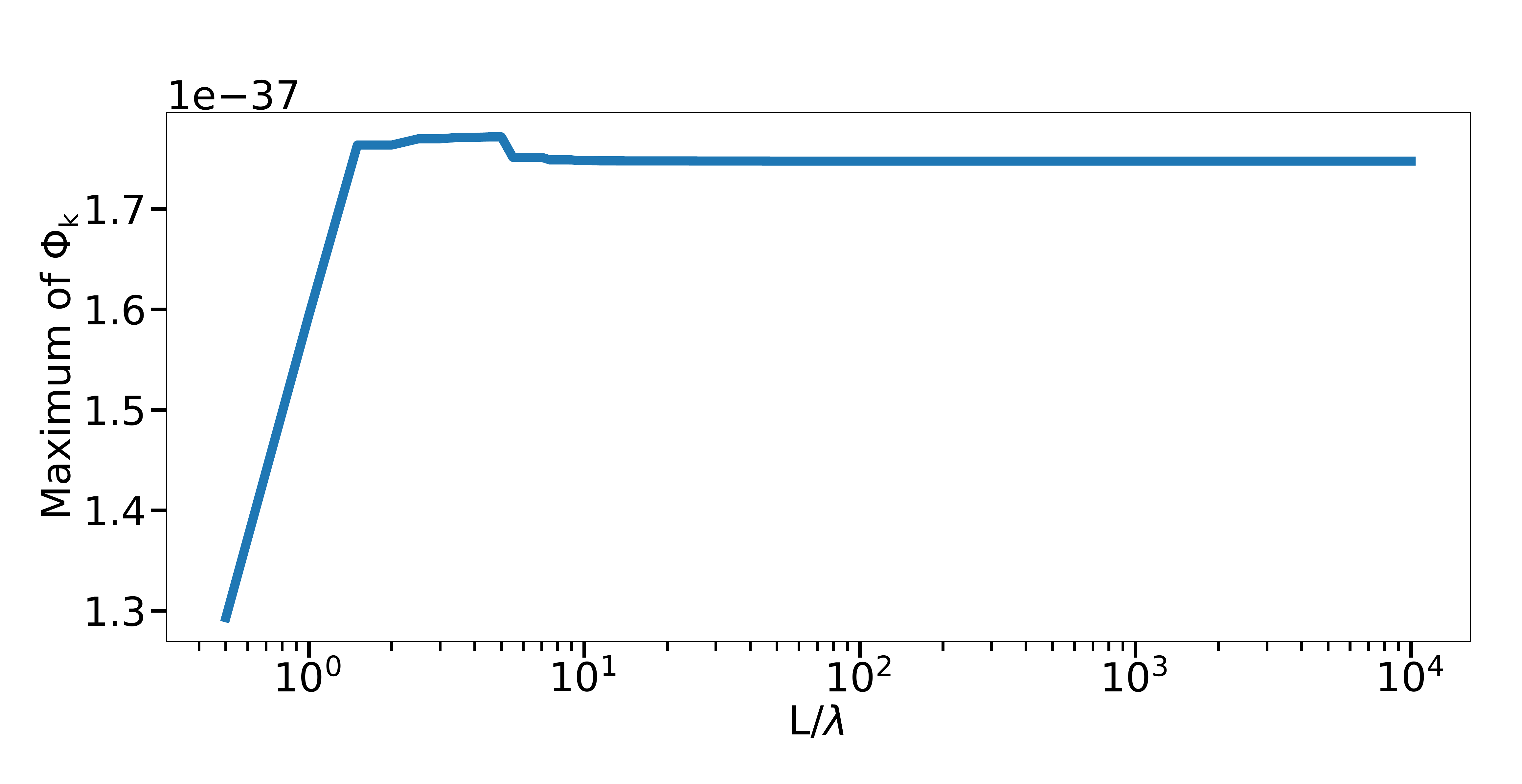}
\caption{Maximum of amplitude of the metric deformation for an oscillating source at constant power $P= 1~\rm PW$ while taking into account the diffraction limit. Results are plotted as a function of the number of optical cycles $L/\lambda$.\label{maxMaxPcteRt}}
\end{center}
\end{figure}

The study of the oscillatory part brought with it the idea of the dependence of the deformation amplitude on the number of optical cycles. We thus want to know if any difference appears on amplitude when we make the former vary. The Figure \ref{maxMaxPcteRt} evaluates this amplitude for $R>L$ and shows us that outside of a fairly negligible increase for the three first number of optical cycles $L/\lambda=0.5,\, 1,\, \text{and} \, 1.5$, the metric perturbation remains constant across all number of optical cycles $L/\lambda$. This result differs from the study on the constant part where the amplitude at fixed power clearly depends logarithmically on this factor, as shown equation \eqref{eq:hZr}.

Having determined the behavior of both the constant and oscillatory part of the source presented equation \eqref{eq:hSource}, we can regroup our findings to determine the metric deformation generated in the longitudinal direction by a linearly polarized beam of light.

\section{Metric perturbation generated by a linearly polarized light pulse}

To explain how exactly a linearly polarized beam of light influences the space-time metric around it, we need to go back to equation \eqref{eq:hSource} which describes the local Einstein equation for the time or $z$ components of the perturbation metric $h_{\mu\nu}$ for a linearly polarized electromagnetic wave.

We thus set the global expression on such a case by taking example of what we did before in equations \eqref{eq:SrcCte} and \eqref{eq:SrcOsc}:
\begin{equation}
\begin{split}
    S(ct,z,r)={\cal{A}}\left[ 1 + \cos(2k(z-ct)) \right] \times \\
    \times H(ct) H(R-r) H(r) H(z-ct) H(L-(z-ct))
\end{split}
\end{equation}

We already know how to solve analytically the Einstein equations for the complete source term, as we know the analytical solution for both the constant part and the oscillatory part of this source term. We can regroup and scale our previous answers as the perturbative Einstein equations are strictly linear.
\begin{equation}
    \phi=\phi_0+\phi_{2k}
\end{equation}
The analytical solution of the Einstein equations for a linearly polarized light pulse is thus the half sum of the analytical solution for a constant source term and for an oscillating term which frequency is the double of the electromagnetic oscillation frequency.

\begin{figure}
\begin{center}
\includegraphics[width=\linewidth]{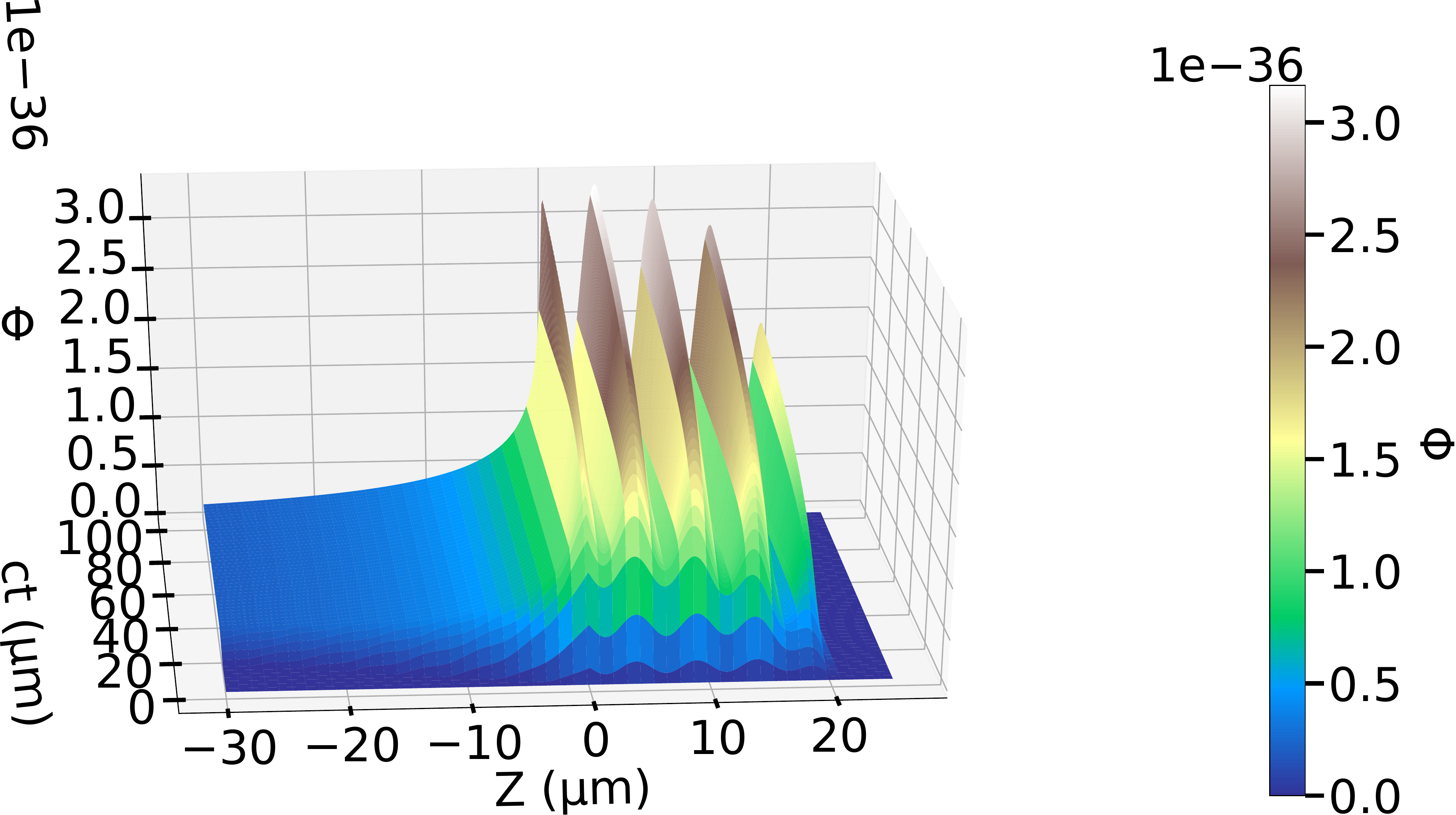}
\caption{3D representation on the comoving plane of the solution for a linearly polarized laser of intensity $I=10^{22}~\rm W/cm^2$ and dimensions of the light cylinder $L=20~\rm~\mu m,~ R=~5\rm~\mu m$.\label{3DTotal}}
\end{center}
\end{figure}

We thus get the following metric deformation profile Figure \ref{3DTotal} for usual laser pulse characteristics $I={\cal{A}}\times c = 10^{22} \rm \, W/cm^2$, $E=525 \rm \, J$, and $R= 5 \, \rm \mu m$, $L=20 \, \rm \mu m$. For readability purposes, we take the electromagnetic oscillation wavelength as $\lambda= 2\pi /k = 10 \, \rm \mu m$.

We now have clearly the sum of both space-time metric deformations we could observe in Figure \ref{Ex3DMob} and \ref{3DOscill}, except that we now get an oscillating metric perturbation at twice the frequency of that of the source electromagnetic wave. We will thus consider again the influence of wavelength and aspect ratio of the source electromagnetic pulse, which is still modeled as a cylinder of electromagnetic energy density.

\begin{figure}
\begin{center}
\includegraphics[width=\linewidth]{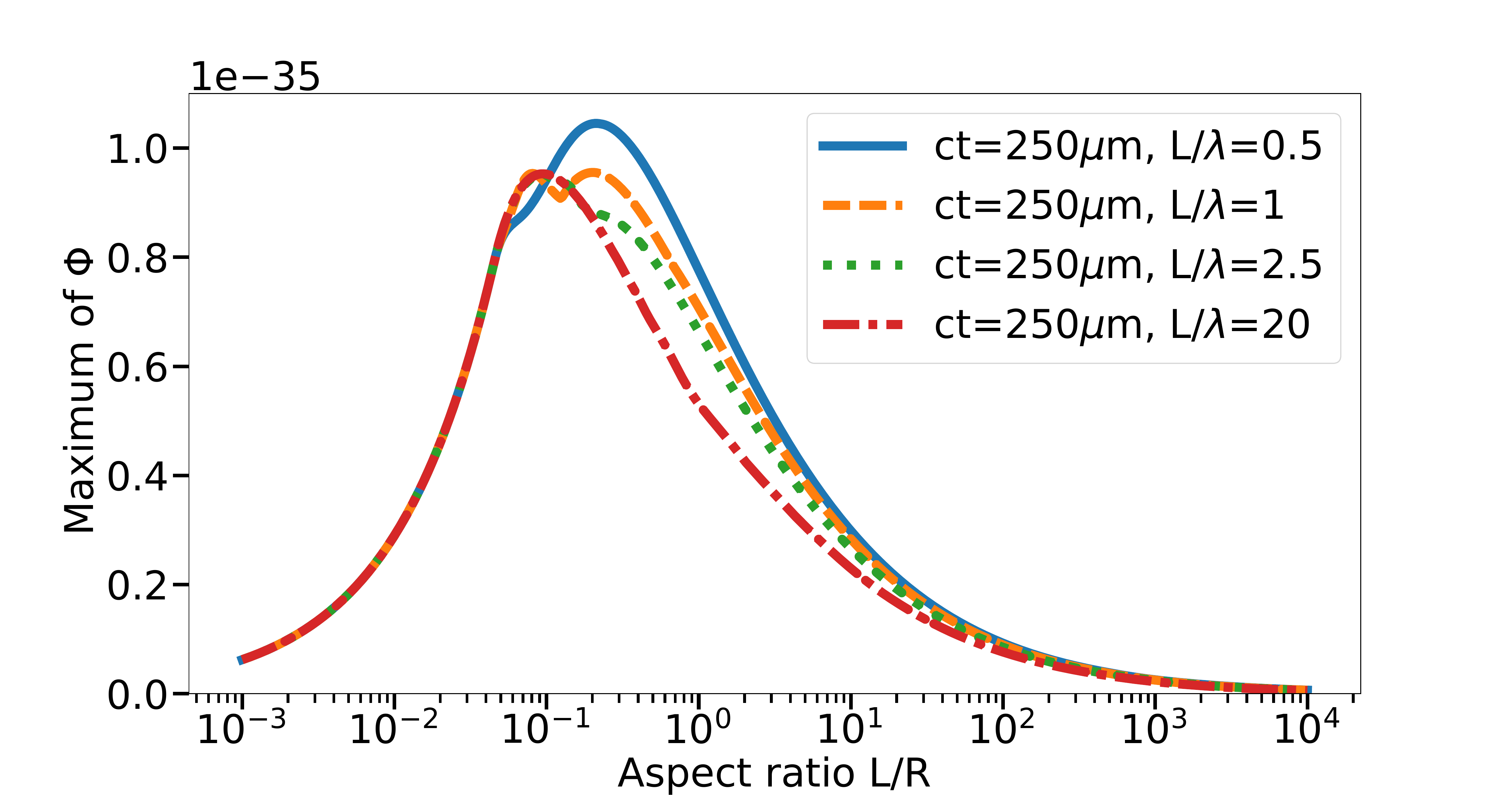}
\caption{Maximum amplitude at fixed time such as $ct=250 \rm \, \mu m$ and a varying number of optical cycles $L/\lambda=0.5$ of the metric perturbation generated by a cylindrical linearly polarized light pulse of intensity $I=2 \times 10^{22} \rm ~W/cm^2$ and energy $E=525 \rm~J$ according to the aspect ratio L/R of said cylinder.\label{maxTotal_k}}
\end{center}
\end{figure}

The profile of the maximum of metric perturbation has a slightly more peculiar shape at low number of electromagnetic oscillations $L/\lambda$, as we can see the contribution of both the constant and oscillatory part. In Figure \ref{maxTotal_k}, two local maxima are indeed distinguishable for the three smaller plotted $L/\lambda$. The first maximum, which is the lowest in aspect ratio, doesn't see its position change, and even doesn't see its amplitude change for the three greater $L/\lambda$. It represents the maximum of deformation introduced by the constant part of the energy density of the electromagnetic source. The local maximum located at greater aspect ratios corresponds to the oscillating part of the electromagnetic source, we can see it behaves like in Figure \ref{maxOscill_k}, as the value of the maximum decreases as the number of oscillations $L/\lambda$ increases. What this seems to suggest is that depending on the aspect ratio and frequency of oscillation of the source electromagnetic energy density, we can observe one of two physical phenomena. Either we get a metric perturbation quite similar to the one presented Figure \ref{3DTotal}, which denotes the gravitational influence of both the constant and oscillatory part of the electromagnetic energy density, or we get a metric perturbation that looks greatly like the deformation at constant energy density presented Figure \ref{Ex3DMob}, with an added comparatively small oscillatory contribution. 

We pursue this study by shifting from the study at constant energy and intensity, to the one at constant power, which we have shown previously to be more fitting to the physical case of a collimated electromagnetic pulse.

\begin{figure}
\begin{center}
\includegraphics[width=\linewidth]{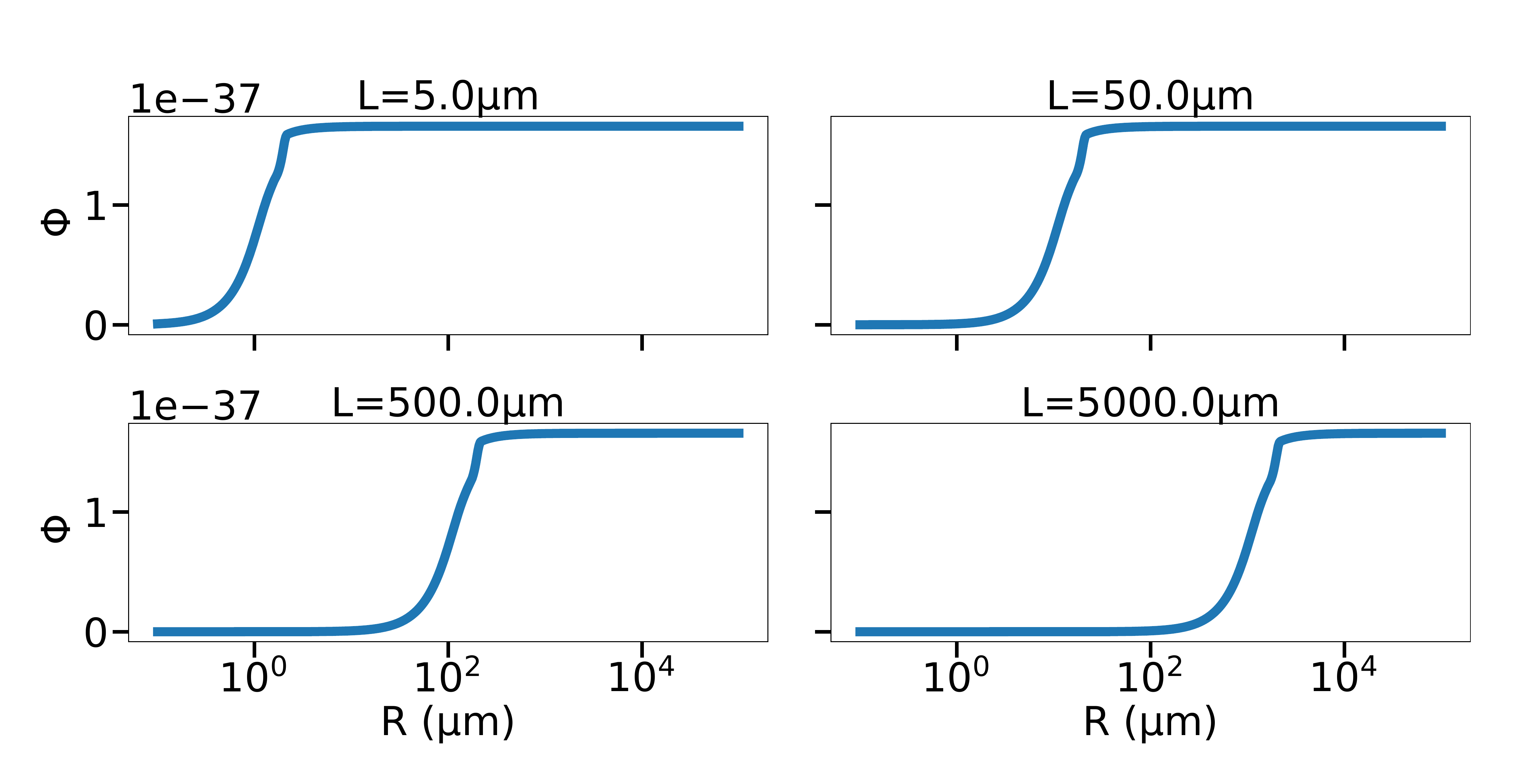}
\caption{Maximum of amplitude at set $L$ {\large (}$L=5 \rm~\mu m$, $L=50~\rm \mu m$, $L=500 \rm~\mu m$, $L=5000 \rm~\mu m${\large )} of the metric deformation at constant power $P= 1~\rm PW$ and number of optical cycles $L/\lambda =0.5$ as a function of the cylinder radius $R$ and at time such that $ct=2~z_R$.\label{maxPcteRtTotal}}
\end{center}
\end{figure}

\begin{figure}
\begin{center}
\includegraphics[width=\linewidth]{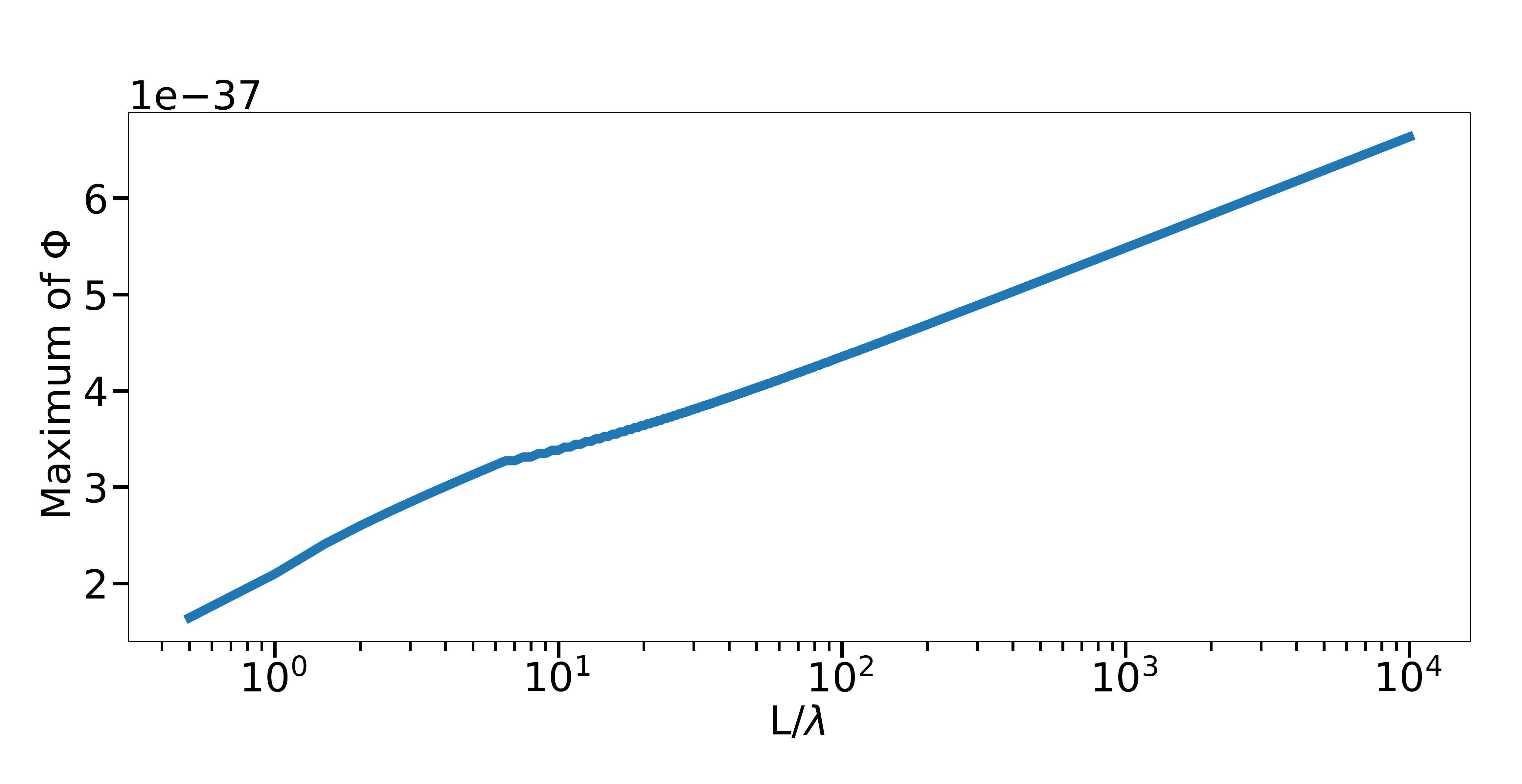}
\caption{Maximum of amplitude of the metric deformation for a linearly polarized light at constant power $P= 1~\rm PW$ while taking into account the diffraction limit. Results are plotted as a function of the number of optical cycles $L/\lambda$.\label{maxMaxPcteRtTotal}}
\end{center}
\end{figure}

Considering the results presented in Figures \ref{maxPcteRt} and \ref{maxPcteRtOscill}, we unsurprisingly find in Figure \ref{maxPcteRtTotal} that for a cylinder radius $R$ of the electromagnetic source large enough before its cylinder length $L$, the metric perturbation generated by a pulse of linearly polarized light is proportional to the source's power. This result is true at a fixed $L/\lambda$ number of oscillation, and we thus need to see, if we still can corroborate the observations made on both Figures \ref{maxTotal_k} and \ref{maxMaxPcteRt}. As such, we make the number of electromagnetic oscillations $L/\lambda$ vary at set power $P=1 \rm \, PW$, and observe the value of the metric deformation thus obtained Figure \ref{maxMaxPcteRtTotal}.

As we can see, we do not obtain a similar result to Figure \ref{maxMaxPcteRt}, as the oscillatory part of the metric deformation does not vary with $L/\lambda$, but the constant part does. Indeed, when increasing $L/\lambda$, we also increase the Rayleigh length $z_R$ on the double of which the electromagnetic source can be considered as cylindrical. As the constant part of the metric deformation does not see its amplitude decrease when $L/\lambda$ increases, it actually, at set power, grows logarithmically with $L/\lambda$ as suggested by the Taylor expansion presented equation \eqref{eq:hZr}.

We thus can, for a pulse of light, set up two different experiments, as it was previously stated for our study at constant energy and intensity. We can, at constant power either observe a case, at low $L/\lambda$, where both a wave shaped and oscillatory part of the metric perturbation is observable, which should be advantageous to observe quick variations of the space-time metric; or we can at high $L/\lambda$ get a more important space-time deformation, while sacrificing the visibility of the oscillatory part of this perturbation, allowing for a more simple, but also smoother shape to be observed, which could be interesting if the goal is to observe directly the perturbation $\phi$ of the space-time metric, and not its variations.

\section{Discussion}

The value of metric deformation $\phi$ found for a beam of power $P=1 \rm~PW$ is of the order of $10^{-37}$, which is still small but shows an improvement compared to the generation of a metric perturbation by matter acceleration. To recall the results of theorized experiments for generation by mass acceleration, the paper by \citet{ribeyre_high_2012} presents the different possible methods of generating metric deformation from mass ablation by high power laser. They give an evaluation of the value of such a deformation for each of the different experiments performed for powers of the order of a PW or energies of the order of a MJ. For these values, they find a deformation of the order of:
\begin{equation}
    \phi_{m}\approx 10^{-40} \sim 10^{-39}
\end{equation}
The main limiting factor to these results is the distance at which the observing device must stand due to the explosive nature of the mass acceleration. \citet{ribeyre_high_2012} position in this paper such a detection device at $10$\,m from the source, a huge distance compared to those considered for the study of the direct generation of a metric deformation by intense light.\\
The results of \citet{kadlecova_gravitational_2017} confirm this evaluation by finding for the studied experiment:
\begin{equation}
    \phi_{m}\approx 5 \times 10^{-40}
\end{equation}
These results shall be put in perspective with the result of this study, which gives us a ratio of at least two orders of magnitude between the deformation generated by mass acceleration and that generated by intense light of:
\begin{equation}
    \dfrac{\phi}{\phi_m}\approx 400
\end{equation}
Note the qualitative difference between the nature of the metric deformation created by mass acceleration compared to that created by a beam of light that we have studied here.

While the first one is of chaotic nature because of the intense and arbitrary nature of the process of ejection and propulsion of mass by the creation of a plasma, the second one is directly generated by a laser, that is to say an object with well determined geometrical properties. \citet{grishchuk_electromagnetic_2003} puts forward in his paper this case of establishment of a coherent gravitational source, compared to a generation by mass acceleration, whose coherence would be difficult to establish and maintain in larger scales.\\
Furthermore, currently observed astrophysical metric deformation are, on one hand, induced by the quadrupolar moment of acceleration for the massive case, which greatly reduce the efficiency of this method, and generates a deformation that is transversal to its propagation, whereas on the other hand the generation by light fully uses the displaced energy and nets us a longitudinal deformation, which constitutes another subtlety that will have to be taken into account in further experiments.

Obtaining a metric deformation of the order of $\phi=10^{-37}$ for a laser of power $P=1\,\rm~PW$ allows us to consider a source of metric deformation which, although still of weak effect, appears to be coherent, easier to observe and especially more adaptable to the frequency range of detection than the generation by matter acceleration.
The perspectives of detection of such a metric deformation remain to be studied. But we can already lift the veil on several interesting detection setups. First, methods of detection of metric perturbations produced by light beams by using high frequency mechanical detectors are detailed in the Ref.~\cite{spengler_perspectives_2021}.
Alternatively, the delay induced on an atomic clock by the repeated exposition to the metric deformation generated by a beam of light could be measured, as the extreme sensibility of atomic clocks to gravitational variations was shown in \cite{mcgrew_atomic_2018}. The related experiment would rely on the capacity of some laser facilities to deliver high power laser pulse at relatively high frequencies, thus allowing to measure the deformation created repeatedly by the accumulation of the induced delay on an atomic clock.

Determining which metric deformation detector would be best fit requires us to clarify what is the frequency of the metric deformation generated inside a pulse of light. Depending on the case we are studying, there are two possibilities for such a frequency. The first one would be the frequency associated with the length of the pulse $L$, such as the metric deformation observed would virtually have a frequency $\nu=c/L$. This frequency is associated with the physical case of a non-oscillating electromagnetic energy-stress tensor, as in a circularly polarised pulse of light. The second one is the frequency of the oscillatory part of the metric deformation, which is associated with the first mentioned frequency in the physical case of a pulse of linearly polarized light. As such, a pulse of linearly polarized light would present two detectable frequencies for a metric deformation detector. Taking as an example a pulse of linearly polarized light of wavelength $\lambda = 800 \, \rm nm$ and length $L= 4 \, \rm \mu m $ at set power $P= 1 \, \rm PW$, we get the two frequencies for metric deformation $\nu_0 = 7.5 \times 10^{14} \, \rm Hz$ and $\nu_{osc}=7.5 \times 10^{15} \, \rm Hz $, for a metric deformation $\phi \sim 10^{-37}$.\\
Detectors at such a frequency band are devised and are expected to currently have a sensitivity up to $10^{-30}$. Such detectors use the inverse Gertsenshtein effect \cite{zeldovich_electromagnetic_1974} and are detailed in the paper of \citet{ejlli_upper_2019}.

Room for improvement can also be made in the laser power, as one could envision lasers who would soon reach a power in the exawatts. Due to the linear dependence of the metric deformation on laser power, this would lead to $\phi\approx 10^{-34}$. This progress could lead to one of the end goal of the generation of oscillating metric deformations in the laboratory, which would be the establishment of the gravitational equivalent to a Hertz experiment, where we would generate and detect a metric perturbation.

Another physical phenomenon that could be of use in proving the validity of such a metric deformation generation would be the Gamma Ray Bursts (GRB). These astronomical phenomena indeed generate extremely powerful light beams $10^{44} - 10^{45}~\rm W$ for a duration up to a whole second, i.e $L=3.0 \times 10^8 \, \rm m$ \citep{piran_physics_2004}. Considering the GRB photons have a relatively uniform energy around $250 \rm~keV$, i.e $\lambda = 50 \rm~nm$, Equation \eqref{eq:hZr} gives us $\phi \approx 10^{-6}$ as an approximate of the amplitude of the metric deformation inside a cylindric enough GRB. This is an impressive result which will probably need its own accurate calculation since, far from its source, a GRB's shape tends to be more conic. This result still gives great hope for the detection of metric deformations induced by light.

\section{Conclusion}

The study of the metric deformation generated by the electromagnetic stress-energy tensor $T^{em}_{\mu,\nu}$ of a light pulse gives us several important results.
First we have put into action a case exhaustion method in order to obtain an analytical solution for the Einstein equations with a source delimited in space. It has been proved to give at long distances the result expected from the \citet{schwarzschild_uber_1916} model for a static cylinder of constant energy density, and gave us more insight on the establishment of such a metric deformation in the frame of the thought experiment of a suddenly appearing amount of energy. This method will prove useful in the calculation of further physical cases, which can also be modelled by an energy density only present in a clearly delimited space. Such cases actually include most compact massive physical objects, as well as some directional radiations.

The metric perturbation generated by a cylinder of constant energy density moving at the speed of light $c$, i.e a circularly polarized pulse of light, presents a singular profile. This wave-like profile along $z$ gives us good hopes for the study of the detection of the deformation by a probe beam, since the light deflection depends on the variation of the metric. Moreover, the study of an electromagnetic pulse of linear polarization doubles down on this advantage by introducing an oscillatory part to the metric, which has double the frequency of the electromagnetic source.

This metric deformation is linearly dependant on the power of the electromagnetic source, as opposed to its intensity. The influence of quantum electrodynamics effects can thus be rendered minimal by not focusing too much the light source.

While the amplitude of the wave-shaped deformation due to the constant part of the energy density tensor is logarithmically dependant on the number of optical cycle $L/\lambda$ of the electromagnetic source, the amplitude of the oscillatory part stays constant. This gives the possibility for future detection experiments to cater the shape of the metric deformation to the tools used, depending on if the oscillatory or wave-shaped part of the deformation is desired to be observed.

Current contenders for the observation of the effect of metric deformation generated by light-only include high power lasers for experiments in laboratory for which the current possible deformation is estimated at $\phi\sim 10^{-37}$, and Gamma Ray Bursts for astrophysical observations, for which the deformation is estimated at $\phi \sim 10^{-6}$.

Further study could concern the influence of such metric deformation on the path and spectrum of a probe light beam, or the more exact calculation of the metric deformation generated by light in extreme astrophysical events.

\section*{Acknowledgements}
This research was supported by the French National Research Agency (Grant No. ANR-17-CE30-0033-01) TULIMA Project and by the NSF (Grants No. 1632777 and No. 1821944) and AFOSR (Grant No. FA9550-17-1-0382). We thank J. L. Dubois, F. Catoire and P. Gonzalez de Alaiza Martinez for the interest they've shown in this study and their advice. We finally thank our reviewers for their helpful input.

	\bibliography{Ma_bibTeX}
	%\printbibliography
 
\end{document}